\documentclass[preprint2]{aastex631}

\usepackage{xfrac}
\usepackage{amsmath}
\usepackage{upgreek}
\usepackage{enumerate}

\newcommand\Rm{{\rm Rm} }
\newcommand\Rey{{\rm Re} }
\newcommand\Pm{{\rm Pm} }

\newcommand\SNr{\dot\sigma_{\rm Sn}}

\newcommand\ESK{E_{\rm kin}}
\newcommand\EST{E_{\rm th}}

\newcommand\Ms{$\cal{M}$}
\newcommand{\vect}[1]{{{\mbox{\boldmath $#1$}}}}
\newcommand{\mathbfss}[1]{\textbf{\textsf{#1}}}
\newcommand\K{~ {\rm K}}
\newcommand\cmcube{~ {\rm cm^{-3}}}
\newcommand\cs{ c_{\rm s}}
\newcommand\cplocal{ c_{\rm p}}

\newcommand\cv{ c_{\rm v}}
\newcommand\fV{ f_{V,i}}
\newcommand\fH{ f_{V,h}}
\newcommand\kpc{~ {\rm kpc}}
\newcommand\pc{~ {\rm pc}}
\newcommand\dx{ {\delta x}}
\newcommand\Myr{~ {\rm Myr}}
\newcommand\Gyr{~ {\rm Gyr}}
\newcommand\erg{~ {\rm erg}}
\newcommand\kms{~ {\rm km~s}^{-1}}

\newcommand\eB{{\langle e_{\rm B}\rangle}}
\newcommand\eK{{\overline{e_{\rm K}}}}
\newcommand\barg{{\overline{\gamma}}}
\newcommand\uOl{\texttt{UP0r10sl}}
\newcommand\hOl{\texttt{HP0r10sl}}
\newcommand\hIl{\texttt{HP1r10sl}}
\newcommand\hIIl{\texttt{HP2r5sl}}
\newcommand\hVm{\texttt{HP5r8sm}}

\newcommand\hVl{\texttt{HP5r2sl}}
\newcommand\mOh{\texttt{MP0r10sh}}
\newcommand\mOIh{\texttt{MP0r5sh}}
\newcommand\mIIl{\texttt{MP2r5sl}}
\newcommand\mOl{\texttt{MP0r10sl}}
\newcommand\lOIh{\texttt{LP0r5sh}}
\newcommand\lOh{\texttt{LP0r10sh}}
\newcommand\lOl{\texttt{LP0r10sl}}
\newcommand\lIIl{\texttt{LP2r5sl}}
\newcommand\lVmS{\texttt{LP5r8sm-S}}

\definecolor{midblue}{rgb}{0.0,0.4,0.7}
\definecolor{midgreen}{rgb}{0.1,0.6,0.3}
\definecolor{mypurple}{rgb}{0.7,0.3,0.8}

\newcommand{\rev}[1]{\textcolor{black}{#1}}

\received{\today}
\revised{\today}
\accepted{}
\submitjournal{ApJ}

\shorttitle{Small-scale dynamo in a multi-phase medium}
\shortauthors{Gent et al.}

\begin{document}

\title{The small-scale dynamo in a multiphase supernova-driven medium}

\correspondingauthor{Frederick Gent}
\email{Email: frederick.gent@aalto.fi, mordecai@amnh.org,\\ maarit.korpi-lagg@aalto.fi, nishant@iucaa.in}

\author[0000-0002-1331-2260]{Frederick A. Gent}
\affiliation{
Astroinformatics, Department of Computer Science, Aalto University, P.O. Box 15400, FI-00076 Espoo, Finland
 }
\affiliation{
    School of Mathematics, Statistics and Physics,
      Newcastle University, NE1 7RU, UK 
 }

\author[0000-0003-0064-4060]{Mordecai-Mark {Mac Low}}
\affiliation{
  Department of Astrophysics, American Museum of Natural History, {79th Street at Central Park West,}
  New York, NY 10024, USA
}
\author[0000-0002-9614-2200]{Maarit J. Korpi-Lagg}
\affiliation{
Astroinformatics, Department of Computer Science, Aalto University, P.O. Box 15400, FI-00076 Espoo, Finland
}
\affiliation{
    Nordic Institute for Theoretical Physics,
      Roslagstullsbacken 23, SE-106 91 Stockholm, Sweden 
}
\author[0000-0001-6097-688X]{Nishant K. Singh}
\affiliation{
Inter-University Centre for Astronomy \& Astrophysics, Post Bag 4, Ganeshkhind, Pune 411 007, India
}


\begin{abstract}
Magnetic fields grow quickly, even at early cosmological times, suggesting the
	action of a small-scale dynamo (SSD) in the interstellar medium (ISM)
	of galaxies.  Many studies have focused on idealized, isotropic,
	homogeneous, turbulent driving of the SSD.  Here we analyze more
	realistic simulations of supernova-driven turbulence to understand how
	it drives an SSD.  We find that SSD growth rates are intermittently
	variable as a result of the evolving multiphase ISM structure. Rapid
	growth in the magnetic field typically occurs in hot gas, with the
	highest overall growth rates occurring when the fractional volume of
	hot gas is large.  SSD growth rates correlate most strongly with
	vorticity and \rev{fluid Reynolds number}, which also \rev{both} correlate \rev{strongly} with gas temperature.  Rotational
	energy exceeds irrotational energy in all phases, but particularly in
	the hot phase while SSD growth is most rapid.  Supernova (SN) rate does
	not significantly affect the ISM average kinetic energy density.
	Rather, higher temperatures associated with high SN rates tend to
	increase SSD growth rates.  SSD saturates with total magnetic energy
	density around 5\% of equipartition to kinetic energy density,
	increasing slightly with magnetic Prandtl number.  While magnetic
	energy density in the hot gas can exceed that of the other phases when
	SSD grows most rapidly, it saturates below 5\% of equipartition with
	kinetic energy in the hot gas, while in the cold gas it attains 100\%.
	Fast, intermittent growth of the magnetic field appears to be a
	characteristic behavior of SN-driven, multiphase turbulence.
\end{abstract}

\section{Introduction}\label{sec:intro}
Magnetic fields pervade the Universe at all scales.  Many astrophysical systems
consist of plasma, in which the highly turbulent motions drive small-scale
dynamos (SSD) that rapidly grow magnetic fluctuations at the scales
characteristic of the turbulence.  Such magnetic fields influence the structure
and dynamics of, for example, star forming molecular clouds
\rev{\citep{MK04,LHF12,FK12,Srid14}} and spiral arms \citep{CF53,BBMSS96,FBSBH11}.
Magnetic fields also grow at larger scales relevant to the shape and structure
of their host, such as the polar fields in stars aligned to the rotational
axis, or galactic fields aligned to the spiral arms or disk
\cite[e.g.,][]{Harnett04,FBSBH11,Beck16}.  Such fields are generated by
large-scale dynamos (LSDs).  To model galactic LSDs self-consistently
\citep[e.g.,][]{GEZR08,HWK09,Gent:2013a} requires its evolving entanglement
with the SSD be included.  The vast separation in scale and growth time makes
this challenging. 

The SSD has been investigated numerically for astrophysically relevant
parameters, such as low magnetic Prandtl number Pm applying to the Sun or stars
\citep[e.g.,][]{SHBCMM05,ISCM07,B11,WKGR22}, high Pm typical of the
interstellar medium (ISM) or intracluster medium
\citep[e.g.,][]{SBK02,Schober12,SBSW20}, and high sonic Mach number \Ms\
\citep[e.g.,][]{Haugen:2004M,FCSBKS11,FSBS14}, which would apply in turbulence
driven by supernova (SN) explosions.

Amplification and decay of magnetic fluctuations in highly compressible fluids
can occur independent of the presence of an SSD
through a process called tangling, where a large-scale field is pushed around
by the turbulent flow, and a fluctuating contribution is generated \citep{RK07,KB16}.
In regions of SN compression, magnetic field strength scales with density, the
exponent of its proportionality depending on the geometry of the compression.
However, observations of the diffuse ISM show no correlation between field
strength and density \citep[e.g.,][]{crutcher2003}.
Observations of galaxies indicate that the turbulent magnetic field strength is
typically larger than that of the large-scale fields. The Solar neighbourhood
random field, for example, is about 1.3 times larger than the local regular
field \citep{BSSW03}.  These fields are roughly in equipartition with the
estimated turbulent kinetic energy density.

It is somewhat uncertain how LSD, tangling and SSD interact and contribute to this
picture. Based on a suite of numerical experiments, \citet{KB16} reported that
while tangling, as expected, is positively correlated with the large scale magnetic
fields, the SSD shows an anti-correlation when the mean component of the magnetic
field becomes strong. Super-equipartition mean fields, which could arise in presence
of fluxes of small scale magnetic helicity, tend to suppress SSD.
Some properties of tangling produced magnetic fluctuations are discussed in
\citet{GMKS21} where it was noted that tangling produces only a linear growth for
a given background mean magnetic field. If the background field
itself grows exponentially, the associated fluctuations by tangling are also
expected to show an exponential growth. However, in the absence of any regular
or mean magnetic field, as is the case in the present work, exponentially
growing solutions for small scale magnetic fields must be attributed to SSD.

The high resolution simulations of \citet{BSB16} include SSD and LSD
simultaneously, but only for isothermal, helically-driven turbulence at $0.1 <
\mbox{ Pm } < 10$.  Galaxy simulations including halo-disk scale flows
\citep[e.g.,][]{RT16,SBADMN19} find SSD but capture no LSD.  Their multiphase
structure is parameterized, not evolved explicitly.  While lower resolution
models of LSD in SN-driven turbulence with galactocentric differential rotation
do not include SSD, \citet{Gent:2013a} appeared to do so in a non-isothermal
model, as confirmed by \citet{GMKS21}.  We do not yet investigate the
interaction of the LSD and SSD here, but examine only the properties of the
SSD. 

\citet{SF22} model SSD in a two-phase ISM (cold and warm) using
large-scale momentum injection to drive the turbulence, rather than
point-source thermal injection as primarily applied here.  In the case of both
solenoidal and compressive forcing they find a large dispersion in the
correlation of magnetic field strength to gas density (see their Figure~5).
Compressive forcing yields a perceptible trend of $B\propto\rho^{0.5}$
for warm gas, and $B\propto\rho^{0.7}$ for cold, consistent with the
relations for compression along field lines and spherically, respectively.

SSD in the cold and warm phases has the same growth rate in \citet{SF22}.
Their cold gas has typically higher \Ms, so, based on their previous isothermal
models \citep{SF21}, SSD should be slower. Overall growth is slower in the
multiphase medium than in an isothermal gas of similar mean \Ms.
\rev{Separation} into cold and warm phases driven by thermal instability
\citep{FGH69, SVG02, BKM07, MSQP12}.
\rev{The \citet{SF21} cooling function has a slightly reduced range of instability than
what we model here.} \rev{Of greater significance is that}
without SN thermal energy injection they also have no hot gas in a quasi-stable
third phase \citep{MO77}.  The models we present here suggest this is a crucial
difference.

The SSD in SN-driven turbulence has been modelled by \citet{BKMM04} and
\citet{GMKS21}.  The second of these included a multi-phase ISM with fractional
volumes of cold, warm and hot gas somewhat consistent with observations. Their
key finding was to confirm that under the conditions prevailing in an ISM
heated by SNe, the SSD is easy to excite and amplifies magnetic fluctuations up
to sub-equipartition levels within a few tens of megayears.  A critical issue
that was not resolved, though, was to explain the erratic growth rate of the
dynamo in these simulations, which was not observed by \citet{BKMM04}.

\begin{figure}
\centering
\includegraphics[trim=0.4cm 0.4cm 0.0cm 0.0cm,clip=true,width=0.48\textwidth]{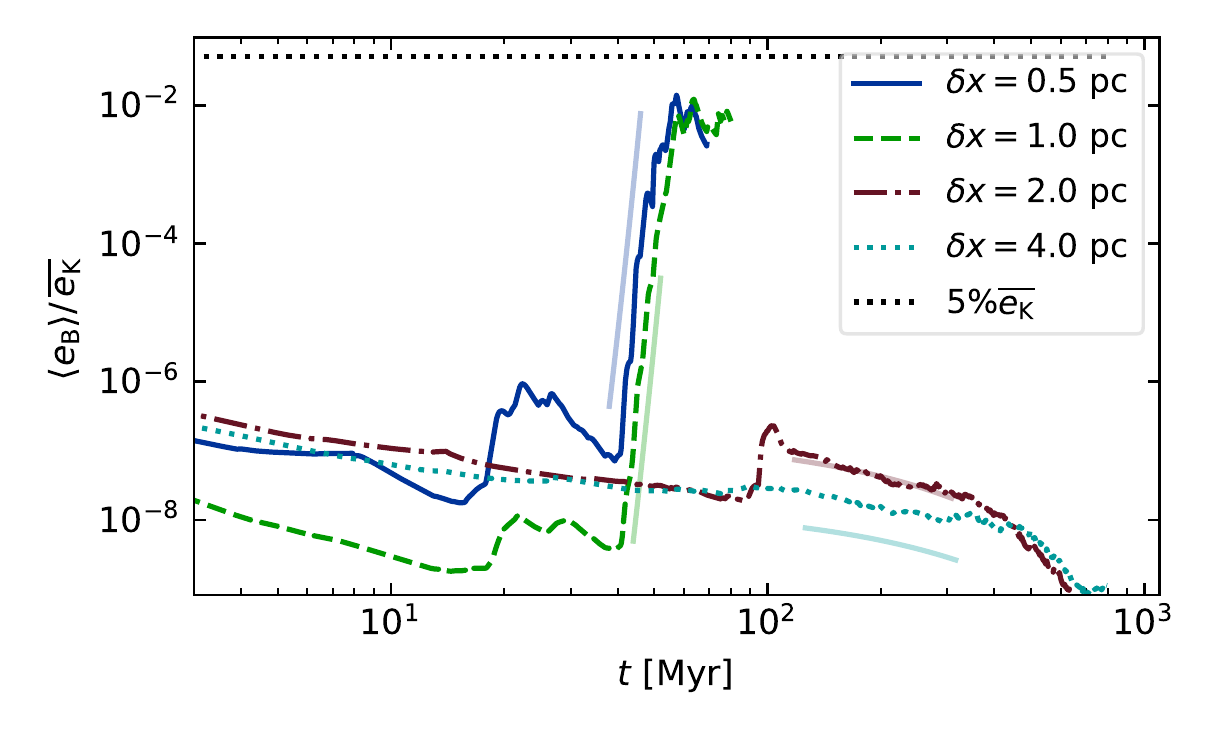}
\includegraphics[trim=0.4cm 0.0cm 0.2cm 0.2cm,clip=true,width=0.48\textwidth]{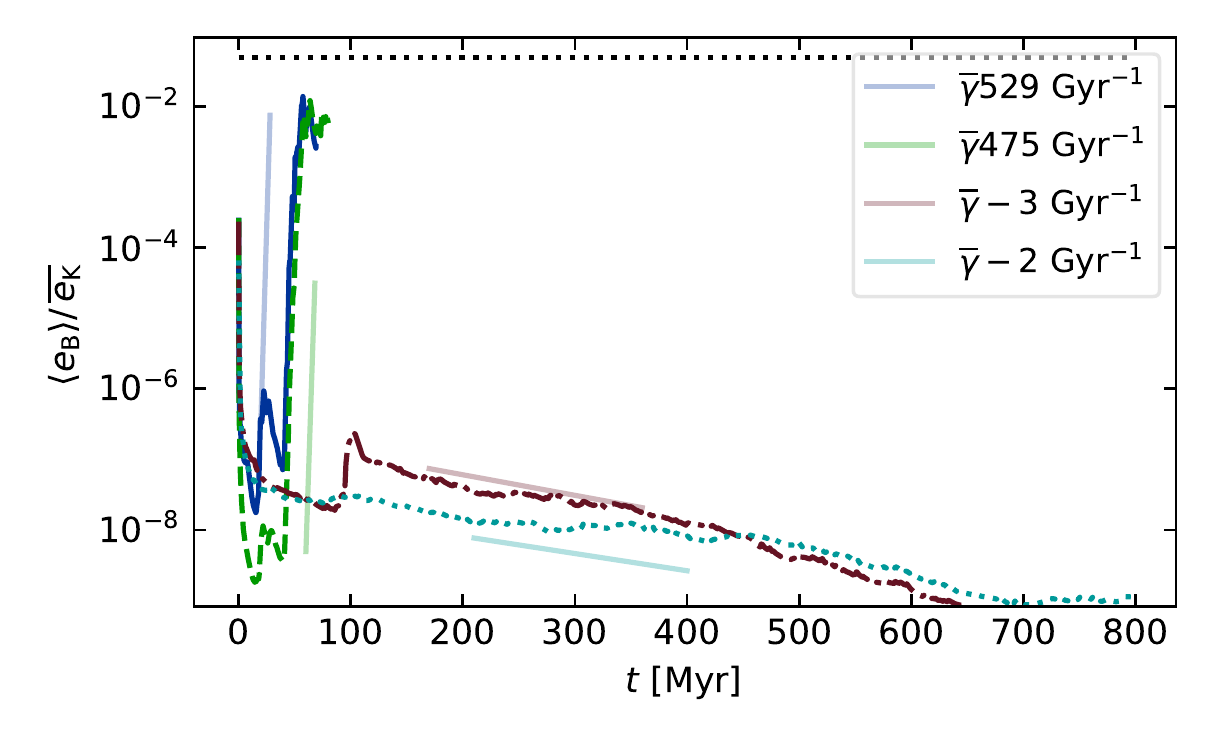}
  \begin{picture}(0,0)(0,0)
    \put(-125,305){{\sf\bf{(a)}}}
    \put(-125,158){{\sf\bf{(b)}}}
  \end{picture}
\caption{
Volume averaged magnetic energy density $\eB$ for Models \uOl, \hOl, \mOl\ and
	\lOl\ (see Table~\ref{tab:models}).  The normalization of $\eB$ is
	given by the time and volume averaged kinetic energy density $\eK$.
	Fits to the exponential growth rate $\barg$, see Eq.\eqref{eq:barg},
	are shown as solid lines of the appropriate color.  Time scale is {\em
	(a)} log, to ease comparison at early times for high resolution, and
	{\em (b)} linear, to show the late time exponential decay of the
	low-resolution models.  For $\delta x\leq1\pc$ the fits span
	$40<t<48\Myr$, while for $\delta x\geq2\pc$ they span $110<t<300\Myr$.
\label{fig:res}
}
\end{figure}

We illustrate this intermittent dynamo in Figure~\ref{fig:res} where we
reproduce data from \citet{GMKS21} Figure~3, but with the inclusion of
least-squares fits of magnetic energy density $\eB\propto\exp(\barg t)$.  False
convergence \citep{FMA91} is apparent at low resolution, although the 2~pc
resolution model does exhibit a brief spell of growth at around 200~megayears.  At
high resolution the solutions are convergent at all times, with slightly more
diffusivity apparent at 1~pc resolution.  Of interest for our study is the
sporadic growth and decay apparent at high resolution between 20 and 40~megayears,
and to some degree also after 50~megayears.  In particular, we seek to understand how
the SSD in the multiphase ISM differs from solutions modelled in isothermal
plasmas, and to explain how this drives such intermittency in SSD behaviour. In
\citet{GMKS21} only the volume averaged growth rates $\barg$ were considered.
In results presented here we include a breakdown by phase.

In Section~\ref{sec:models} we explain the motivation and design of our
experiments.  In Section~\ref{sec:results} we report on the key drivers that
change SSD growth within and between models.  This includes consideration of
how magnetic Prandtl number varies in the multiphase environment of the ISM.
We conclude with a discussion of the significance of these results in
Section~\ref{sec:conc}.

\begin{deluxetable*}{lccccccccccccccr}
\tablecaption{Models included.
Prefixes `U', `H', `M', `L' refer, respectively, to ultra, high, medium and low
	resolution.  Numbers after `P' indicate the nominal magnetic Prandtl
	number $\nu/\eta$.  The `r' numbers indicate the resistivity
	coefficient $\eta$ in units of $10^{-4}$ kpc km s$^{-1}$.  The
	supernova rate `sh', `sm' and `sl', denote 1, 0.5 and 0.2 times the
	Solar neighbourhood rate, respectively. `S' denotes the stratified
	model.  All models use coefficients 2, 5 and 2 for shock handling terms
	with $\zeta_D,~\zeta_\nu$ and $\zeta_\chi$,
	respectively.\label{tab:models}
\tablewidth{\paperwidth}
}
	\tablehead{
Model &&& resolution&&& $\dot\sigma$&&& $\eta$          &&&$\nu$             &&& $\eta_3,\nu_3$\\
      &&& [pc]      &&&[$\SNr$]     &&&[kpc km s$^{-1}$]&&& [kpc km s$^{-1}$]&&&}
\startdata                                                                    
\uOl  &&& 0.5       &&&  0.2        &&& $1(-3)$         &&&    0.0           &&& 1.6 $(-16)$   \\[0.15cm]
\hOl  &&& 1.0       &&&  0.2        &&& $1(-3)$         &&&    0.0           &&& 3.5 $(-15)$   \\
\hIl  &&& 1.0       &&&  0.2        &&& $1(-3)$         &&& $1(-3)$          &&& 2.5 $(-15)$   \\
\hIIl &&& 1.0       &&&  0.2        &&& $5(-4)$         &&& $1(-3)$          &&& 2.5 $(-15)$   \\
\hVl  &&& 1.0       &&&  0.2        &&& $2(-4)$         &&& $1(-3)$          &&& 2.5 $(-15)$   \\
\hVm  &&& 1.0       &&&  0.5        &&& $8(-4)$         &&& $4(-3)$          &&& 2   $(-15)$   \\[0.15cm]
\mOl  &&& 2.0       &&&  0.2        &&& $1(-3)$         &&&    0.0           &&& 8.25$(-14)$   \\
\mIIl &&& 2.0       &&&  0.2        &&& $5(-4)$         &&& $1(-3)$          &&& 8.25$(-14)$   \\
\mOh  &&& 2.0       &&&  1.0        &&& $1(-3)$         &&&    0.0           &&& 8.25$(-14)$   \\
\mOIh &&& 2.0       &&&  1.0        &&& $5(-4)$         &&&    0.0           &&& 8.25$(-14)$   \\[0.15cm]
\lOl  &&& 4.0       &&&  0.2        &&& $1(-3)$         &&&    0.0           &&& 2   $(-12)$   \\
\lIIl &&& 4.0       &&&  0.2        &&& $5(-4)$         &&& $1(-3)$          &&& 2   $(-12)$   \\
\lVmS &&& 4.0       &&&  0.5        &&& $8(-4)$         &&& $4(-3)$          &&& 1   $(-11)$   \\ 
\lOh  &&& 4.0       &&&  1.0        &&& $1(-3)$         &&&    0.0           &&& 2   $(-12)$   \\
\lOIh &&& 4.0       &&&  1.0        &&& $5(-4)$         &&&    0.0           &&& 2   $(-12)$   \\
\enddata
\end{deluxetable*}

\section{Method} \label{sec:models}

\subsection{MHD equations}\label{sec:eq}
We use the Pencil Code
\citep{brandenburg2002,Pencil-JOSS} to model SN-driven turbulence as
described previously in \citet{Gent:2012}, \citet{Gent:2013b}, and \citet{GMKSH20}.  We
solve the system of nonideal, compressible, nonadiabatic, MHD equations
  \begin{eqnarray}
  \label{eq:mass}
    \frac{D\rho}{Dt} &=& 
    -\rho \nabla \cdot \vect{u}
    +\nabla \cdot\zeta_D\nabla\rho,\\[0.5cm]
  \label{eq:mom}
    \rho\frac{D\vect{u}}{Dt} &=& 
    \nabla{\ESK\sigma}
    -\rho~\cs^2\nabla\left({s}/{\cplocal}+\ln\rho\right)
    +\vect{j}\times\vect{B}
    \nonumber\\
    &+&\nabla\cdot \left(2\rho\nu{\mathbfss W}\right)
    +\rho~\nabla\left(\zeta_{\nu}\nabla \cdot \vect{u} \right)
    \nonumber\\
    &+&\nabla\cdot \left(2\rho\nu_6{\mathbfss W}^{(6)}\right)
  {-\vect u\vect{\nabla}\cdot\left(\zeta_D\vect{\nabla}\rho\right)},\\[0.5cm]
  \label{eq:ind}
    \frac{\partial \vect{A}}{\partial t} &=&
    \vect{u}\times\vect{B}
    +\eta\nabla^2\vect{A}
    +\eta_6\nabla^6\vect{A},\\[0.5cm]
  \label{eq:ent}
    \rho T\frac{D s}{Dt} &=&
    \EST\dot\sigma +\rho\Gamma_{\rm UV}
    -\rho^2\Lambda +\eta\mu_0\vect{j}^2 
    \nonumber\\
    &+&2 \rho \nu\left|{\mathbfss W}\right|^{2}
    +\rho~\zeta_{\nu}\left(\nabla \cdot \vect{u} \right)^2
    \nonumber\\
    &+&\nabla\cdot\left(\zeta_\chi\rho T\nabla s\right)
    +\rho T\chi_6\nabla^6 s
    \nonumber\\
    &-& \cv~T \left(
    \zeta_D\nabla^2\rho + \nabla\zeta_D\cdot\nabla\rho\right),
  \end{eqnarray}
with the ideal gas equation of state closing the system, assuming an adiabatic
index (ratio of specific heats) $\cplocal/\cv=5/3$.  Most variables take their
usual meanings; a list of notations used is given in Table~\ref{tab:notation}.

\begin{table*}[h]\caption{\label{tab:notation} Meanings of variables}
\begin{tabular}{>{\centering\arraybackslash}p{1.6cm}>{\centering\arraybackslash}p{6.2cm}>{\arraybackslash}p{8.5cm}}
\hline\hline
{Symbol} & {Denoting} & {Units/Definition}\\\hline
 $e_{\rm B}$ & magnetic energy density & [erg cm$^{-3}$]\\
 $\eK$ & time-averaged kinetic energy density & [erg cm$^{-3}$]\\
 $\barg$ & volume-averaged $\eB$ growth rate & [Gyr$^{-1}$]\\
 $\Gamma$ & relative $e_{\rm B}$ growth rate & [Gyr$^{-1}$]\\
 $\tilde\Gamma$ & absolute $e_{\rm B}$ growth rate & [Gyr$^{-1}$]\\
 $\dot\sigma$ & SN explosion rate & [kpc$^{-3}$ Myr$^{-1}$]\\
 $\dfrac{D}{Dt}$ & material derivative & $\dfrac{\partial }{\partial t}+\vect{u}\cdot \nabla$ \\
 $\nabla$ & gradient vector & e.g., $\left(\dfrac{\partial }{\partial x},\dfrac{\partial }{\partial y},\dfrac{\partial }{\partial z}\right)$ \\
 $\rho$ & gas density & [g $\cmcube$]  \\
 $\vect u$ & gas velocity & [km s$^{-1}$] \\
 $t$ & time & [Myr] \\
 $s$ & specific entropy & [erg g$^{-1}$ K$^{-1}$] \\
 $T$ & gas temperature & [K] \\
 $\vect A$ & magnetic vector potential & [G cm] \\
 $\vect B$ & magnetic field & [G] \\
 $\vect j$ & current density & [erg cm$^{-4}$ G$^{-1}$] \\
 $\mathbfss W$ & traceless rate of strain tensor &
   ${\mathsf W}_{ij} = \dfrac{1}{2}\left(\dfrac{\partial u_i}{\partial x_j}
                  + \dfrac{\partial u_j}{\partial x_i}
                  -\dfrac{2}{3} \delta_{ij}\nabla\cdot \vect u\right)$ \\
 $|\mathbfss W|^2$ & contraction of $\mathbfss W$ &
   $|\mathbfss W|^2={\mathsf W}_{ij}{\mathsf W}_{ij}$\\
 $\mathbfss W^{(6)}$ & 6th order rate of strain tensor &
   ${\mathsf W}_{ij}^{(6)} = \dfrac{1}{2}\left(\dfrac{\partial^5 u_j}{\partial x_i^5}
                  + \dfrac{\partial^4}{\partial x_i^4}\left(\dfrac{\partial u_i}{\partial x_j}\right)
                  -\dfrac{1}{3}\dfrac{\partial^4}{\partial x_i^4}\left(\nabla\cdot \vect u\right)\right)$ \\
 $\zeta_{D},\zeta_{\nu},\zeta_{\chi}$ & shock diffusion coefficients& $\propto \left(-\nabla\cdot\vect u\right)_+^2$\\
 $\nu,\eta$ & viscosity, resistivity coefficients& [kpc km s$^{-1}$]\\
 $\nu_6,\chi_6,\eta_6$ & hyperdiffusion coefficients& [kpc$^{5}$ km s$^{-1}$]\\
 $\Gamma_{\rm UV}$ & UV-heating& [erg g$^{-1}$ s$^{-1}$]\\
 $\Lambda$ & radiative cooling& [erg cm$^{3}$ g$^{-2}$ s$^{-1}$]\\
 $\ESK+\EST$ & SN explosion energy& [10$^{51}$erg]\\
 $\mu_0$ & vacuum magnetic permeability & 1 \\
 $\cs$ & sound speed & [km s$^{-1}$]\\
 $\cplocal$ & specific heat at constant pressure & [erg g$^{-1}$ K$^{-1}$]\\
 $\cv$ & specific heat at constant volume   & [erg g$^{-1}$ K$^{-1}$]\\
\hline
\end{tabular}
\end{table*}

Terms containing $\nu_6,~\chi_6$ and $\eta_6$ apply sixth-order hyperdiffusion
to resolve grid-scale instabilities \citep[see, e.g.,][]{ABGS02,HB04}, with
mesh Reynolds number set to be $\simeq1$ at the scale of the zone size $\dx$.
Terms including $\zeta_D,~\zeta_\nu$ and $\zeta_\chi$ resolve shock
discontinuities with artificial diffusion of mass, momentum, and thermal
energy, respectively.  They depend quadratically on the local strength of the
shock \citep[see][for details]{GMKSH20}.  Equations~\eqref{eq:mom} and
\eqref{eq:ent} include momentum and energy conserving corrections for the
artificial mass diffusion $\zeta_D$  applying in Equation~\eqref{eq:mass}.
Following \citet{GMKS21}, resistive shock diffusion is omitted from
Eq.~\eqref{eq:ind}.

SNe at rate $\dot\sigma$ inject $\EST = 10^{51}\erg$ thermal energy.  In dense
regions up to 5\% is instead injected as kinetic energy $\ESK$
\citep[see][]{KO15,GMKSH20}.  Nonadiabatic heating $\Gamma_{\rm UV}$ and
cooling $\Lambda (T)$ \citep[as detailed in][]{Gent:2013b} follow
\citet{Wolfire:1995} and \citet{Sarazin:1987}.

\subsection{Model parameters}\label{sec:par}

We simulate a nonstratified, nonrotating domain with initial gas number density
$n=1\cmcube$ covering (256 pc)$^3$ and periodic across all boundaries.
Resolution along each edge spans 0.5 to $4\pc$, corresponding to grid sizes of
$64^3$ to $512^3$ zones. Each model SN rate is given relative to the estimated
rate in the Solar neighbourhood of the Milky Way, $\SNr=50\Myr^{-1}\kpc^{-3}$.
SNe occur uniform randomly in space at times following a Poisson process.
Models with common $\dot\sigma$ use the same SN schedule and locations.  For
ambient gas number density $n\simeq1\cmcube$ and $\dot\sigma\simeq\SNr$ an
estimated forcing scale of 60--100 pc \citep{joung2006, avillez2007, HSSFG17}
should support at least 2--4 turbulent cells.  For higher SN rates the forcing
scale reduces, increasing the number of turbulent cells \citep{JMB09}, but the
forcing scale remains unchanged for the lower SN rates applying here, where an
individual SN at gas number density $1\cmcube$ merges with the local sound
speed within a 70 pc radius \citep{GMKSH20}  

In addition, Model \lVmS, is vertically stratified without rotation and has
open vertical boundaries. Initially $n\simeq1\cmcube$ at the midplane.  With
this we examine the effect that vertical expansion of SN remnants or
superbubbles has on the SSD.  Here $\Gamma_{\rm UV}$ is amplified to
$3.5\times$ that of \citet{Wolfire:1995} to support the thickness of the disk
in the absence of ionization heating and cosmic ray pressure gradients
\citep[see][]{HMGI18}.  The domain size of this model is $512\pc$ along the
disk and $\pm 1.534\kpc$ perpendicular to it.  SN are located uniform random
horizontally and normal random with scale heights of $90\pc$ and $325\pc$ for
Type I and Type II SNe, respectively \citep{Ferriere01}. This model otherwise
is as described in \citet{Gent:2013b,Gent:2013a}.

All models are listed in Table~\ref{tab:models} where the model labelling
convention is explained.

 \subsection{Averaging conventions}
Angular brackets indicate the quantity is averaged over the volume, or also
with a subscript $T$ when over \rev{the volume of} individual temperature phases.  An overbar
indicates averaging over a domain {\em and} an interval of time, intervals
which may vary, as explained in the text.

In the case of $\eK$ the interval is selected for each model to exclude initial
transients and the period after the SSD approaches saturation.  Even with
saturation around 5\% of equipartition energy, some models show damping in
kinetic energy. Models with identical SN rates, have near identical kinetic
energy density evolution, so direct comparisons are not affected by this
choice. The magnetic energy density $\langle e_{\rm B}\rangle$ is then
normalized by $\eK$ to ease model comparison.

\subsection{Growth rates and Reynolds number}
The erratic growth of the volume averaged magnetic energy shown in
Figure~\ref{fig:res} and discussed in Sect.~\ref{sec:results} shows that the
kinematic stage of the SSD does not have a well-defined single growth rate
\citep{GMKS21}.  SSD growth typically follows an exponential of the form
\begin{equation}\label{eq:barg}
\eB(t)=\eB(t_0)\exp[\barg(t-t_0)].
\end{equation}
In Figure~\ref{fig:res} we fit such a function to specified time intervals for
each model, but the SSD growth (or decay) at other times differs, with $\barg$
varying or no clear exponential behaviour occurring. Indeed within the domain,
we can find very different growth patterns within and between thermal phases.

To interpret how the time and volume averaged growth rate $\barg$ is influenced
by the phases and dynamical properties, we can use Eq.~\eqref{eq:ind} to
identify instantaneous changes to the magnetic energy. Hyperdiffusion, being
purely numerical, is neglected. Taking the curl of Eq.~\eqref{eq:ind} and
contracting it with $\vect{B}\mu_0^{-1}$, we obtain an equation for the change
of the magnetic energy density
  \begin{eqnarray}
  \label{eq:eB}
    \frac{\partial e_{\rm B}}{\partial t} &=&
    \vect{B}\cdot\nabla\times\left(\vect{u}\times\vect{B}
    +\eta\nabla^2\vect{A}\right)\mu_0^{-1}.
  \end{eqnarray}
Negative values represent decay of the magnetic field and positive its
amplification.  Dividing by $e_{\rm B}$, we obtain an equation for the {\em
relative} growth rate exponent, of the form $e_{\rm B} \propto \exp(\Gamma t)$,
at each instant in time and location in space
  \begin{eqnarray}
  \label{eq:logeB}
    \Gamma (\vect{x},t)  &=&\frac{\vect{B}\cdot\nabla\times\left(\vect{u}\times\vect{B}
    +\eta\nabla^2\vect{A}\right)}{e_{\rm B}\mu_0}.
  \end{eqnarray}
Here, $\Gamma (\vect{x},t)$ is a function of position \vect{x}, distinct from
$\barg$ defined above, which is the volume and time averaged quantity.
Statistical analysis of $\Gamma$ in relation to various physical properties
will help determine how the varying growth depends on the multiphase structure
of the ISM.

The growth rate $\Gamma$ does not indicate the absolute magnitude of the energy
change.  We also need to identify where the largest changes in magnetic energy
density occur, since these need not correlate with the growth rates. We
therefore also define the {\em absolute} growth rate
  \begin{eqnarray}
  \label{eq:tildeG}
    \tilde\Gamma (\vect{x},t)  &=&
    \frac{\vect{B}\cdot\nabla\times\left(\vect{u}\times\vect{B}
    +\eta\nabla^2\vect{A}\right)}{\eB(t)\mu_0},
  \end{eqnarray}
replacing $e_{\rm B}$ in the denominator with the time dependent volume
averaged $\eB(t)$.  Rescaling by time dependent $\langle e_{\rm B}\rangle$
assists comparison between all stages of the SSD.  Both $\Gamma$ and
$\tilde\Gamma$ have values that span orders of magnitude of both signs. To take
advantage of logarithmic scales in our histograms we omit negative and
negligible growth rates.

Similarly, a field of values for the magnetic Reynolds number Rm is calculated
directly from the induction equation by taking the ratio of the advection to
the diffusion terms,
\begin{eqnarray}
  \label{eq:Rm}
    \Rm(\vect{x},t) &=&
    \dfrac{|\nabla\times\vect{u}\times\vect{B}|}{|\eta\mu_0\nabla\times\vect{j}|},
\end{eqnarray}
where $\vect{j}=\mu_0^{-1}\nabla\times\vect{B}$.  In a related approach,
\citet{EGSFB17} decompose the terms of Equation~\eqref{eq:mom} to identify
separately the spatial variation of each force.

\begin{table}
\caption{
Data ranges for sampling.
\label{tab:sampling}}
\begin{tabular}{lc}
\hline
\hline
phase & temperature range \\
\hline
cold & $T<3000\K$\\
warm &$3000\leq T<5\cdot10^4\K$\\
hot  &$T>2.5\cdot10^5\K$ \\\\
\hline
$\delta x$ & time interval \\
\hline
0.5 \& 1 pc & $15<t<55\Myr$ \\
2 pc        & $50<t<150\Myr$ \\
4 pc        & $50<t<300\Myr$ \\
\hline
\end{tabular}
\end{table}

\subsection{Phase fractional volume}

There are various approaches to measuring the proportion of the ISM that
contains gas in each phase, as discussed in detail within \citet{Gent:2013b}.
Here, we measure the fractional volume of SSD activity for each phase $i$
defined as 
\begin{equation}\label{eq:fV}
 \fV = \frac{V_i}{V}, 
\end{equation}
where $V_i$ is the volume occupied by the gas in the temperature range defining
phase $i$, as listed in Table~\ref{tab:sampling}, and $V$ is the total volume.
Both the phase volume and the total volume omit locations with negative or
negligible $\Gamma$ to focus on SSD growth.  For each phase, $\fV$ is computed
from the snapshot within the time intervals listed in Table~\ref{tab:sampling}. 

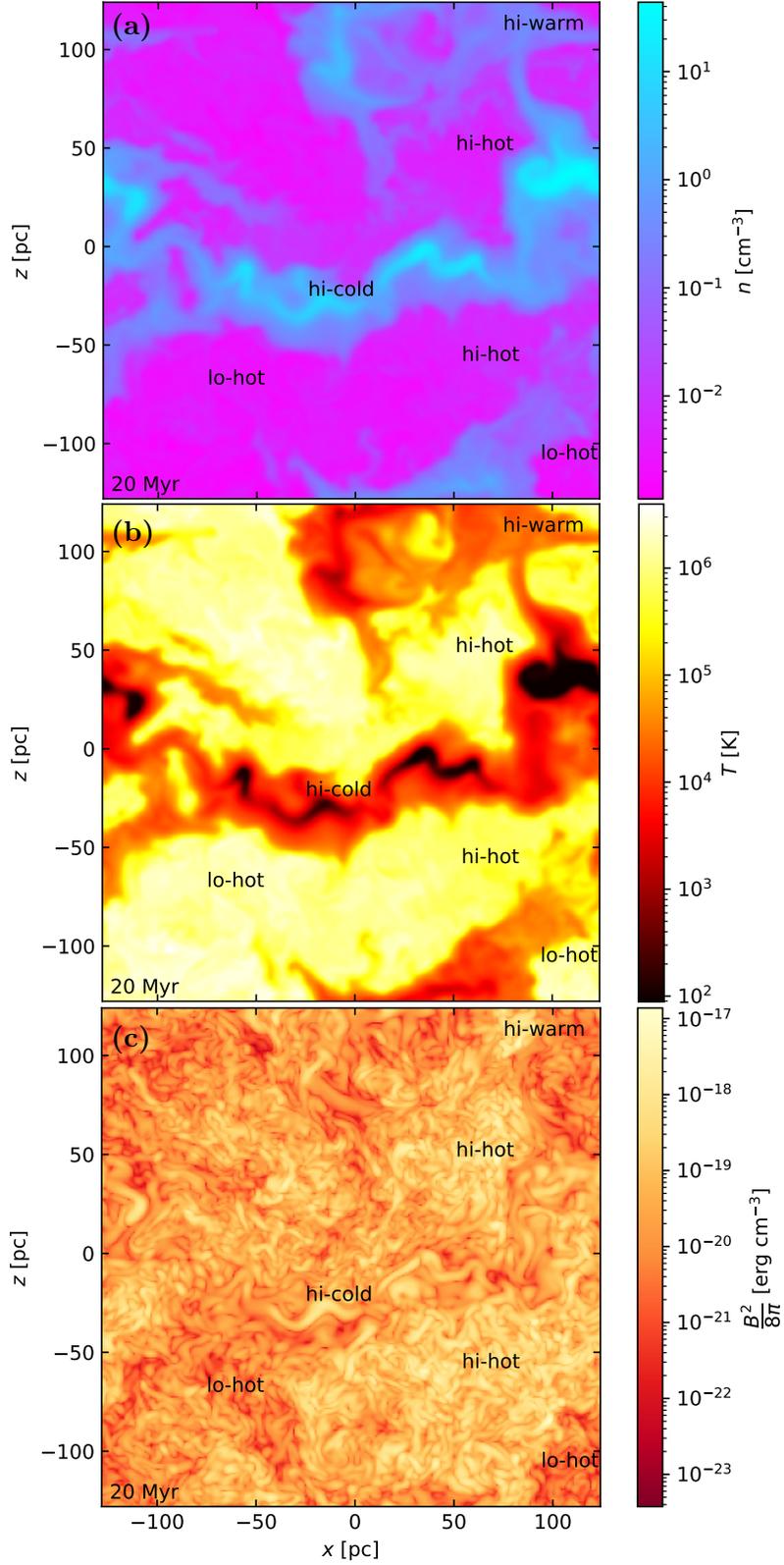
\begin{figure*}
\centering
\includegraphics[trim=1.09cm 1.42cm -0.05cm 0.37cm,clip=true,width=0.600\textwidth]{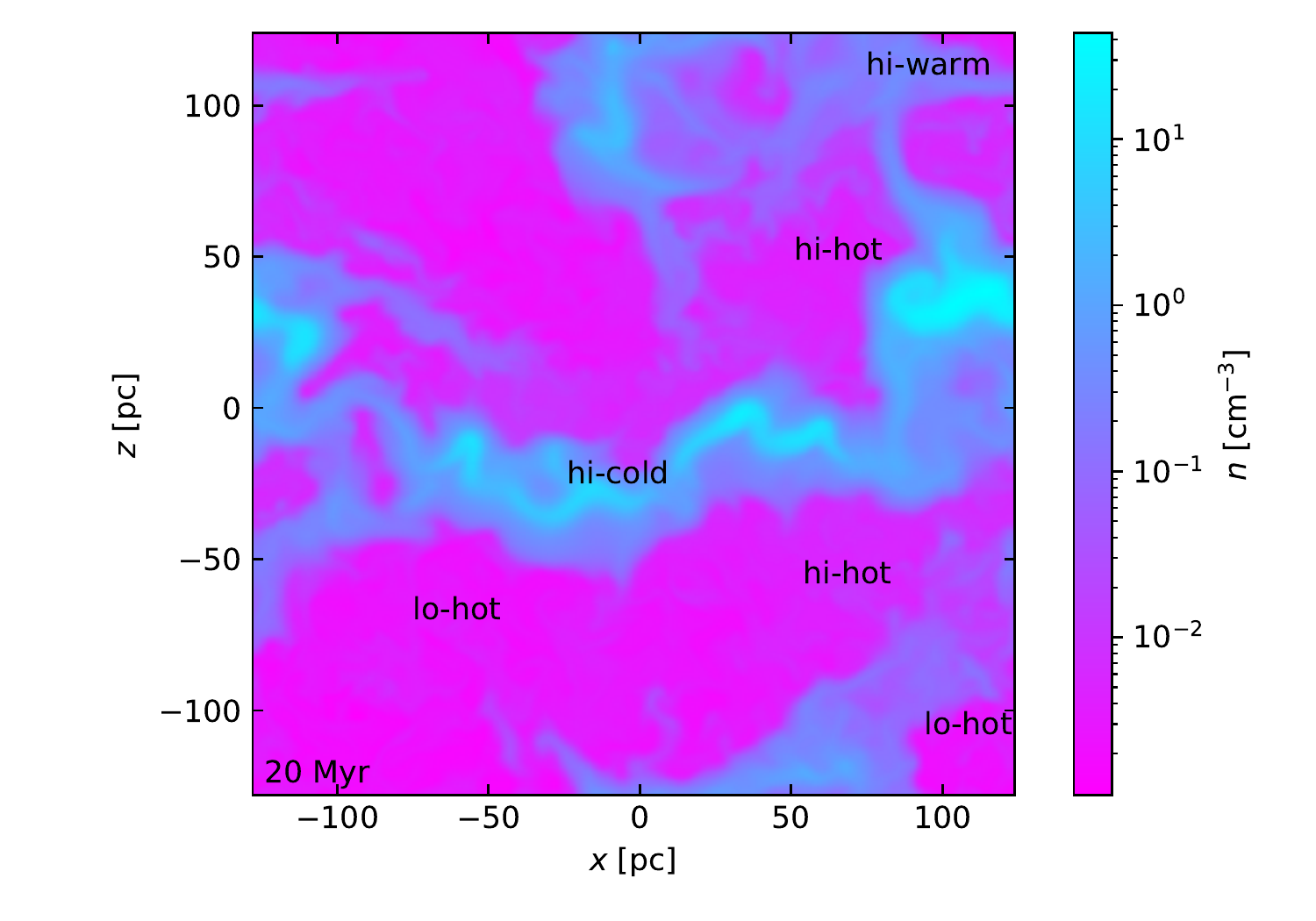}
\includegraphics[trim=1.34cm 1.42cm -0.28cm 0.37cm,clip=true,width=0.600\textwidth]{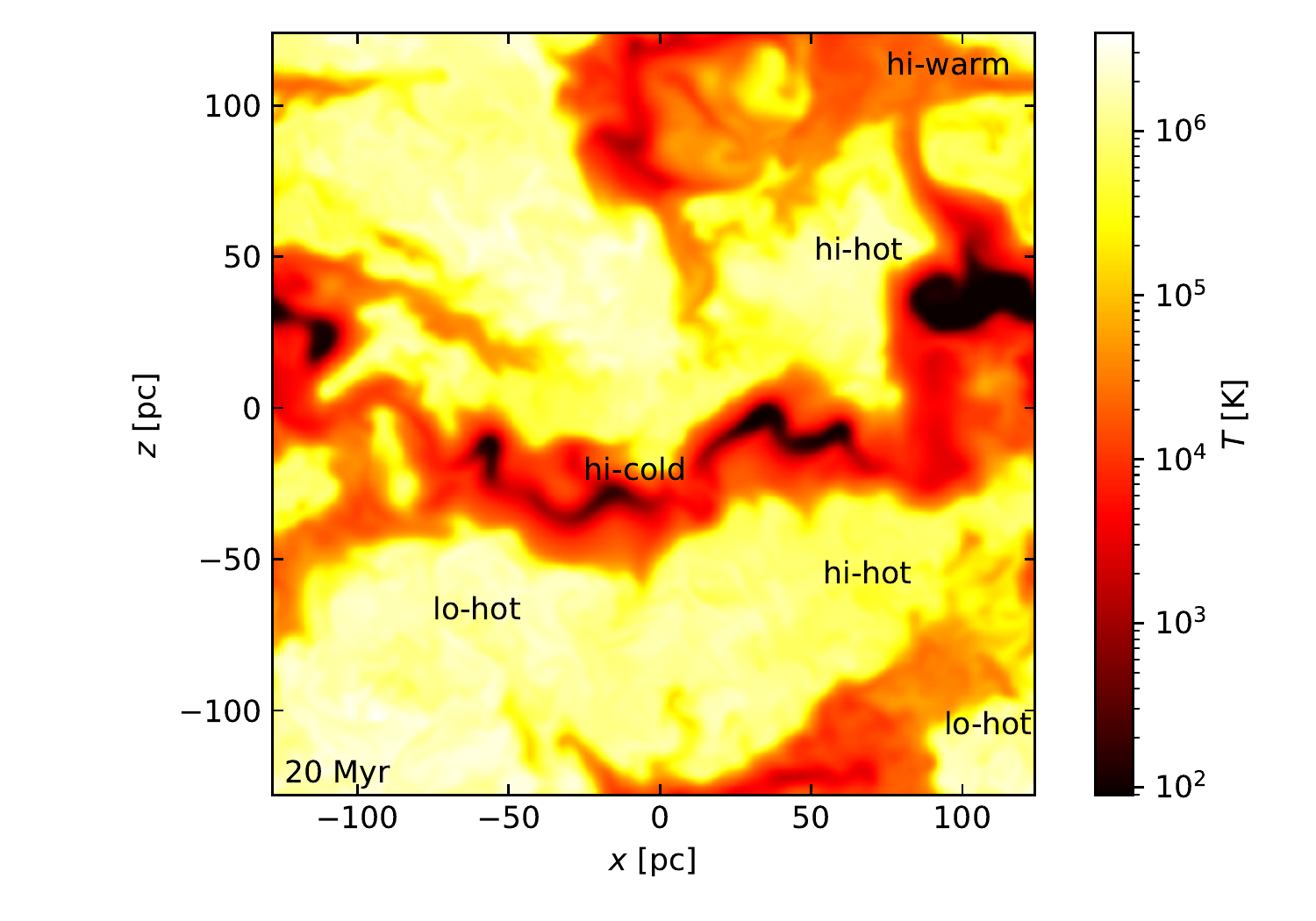}
\includegraphics[trim=0.59cm 0.43cm  0.55cm 0.37cm,clip=true,width=0.600\textwidth]{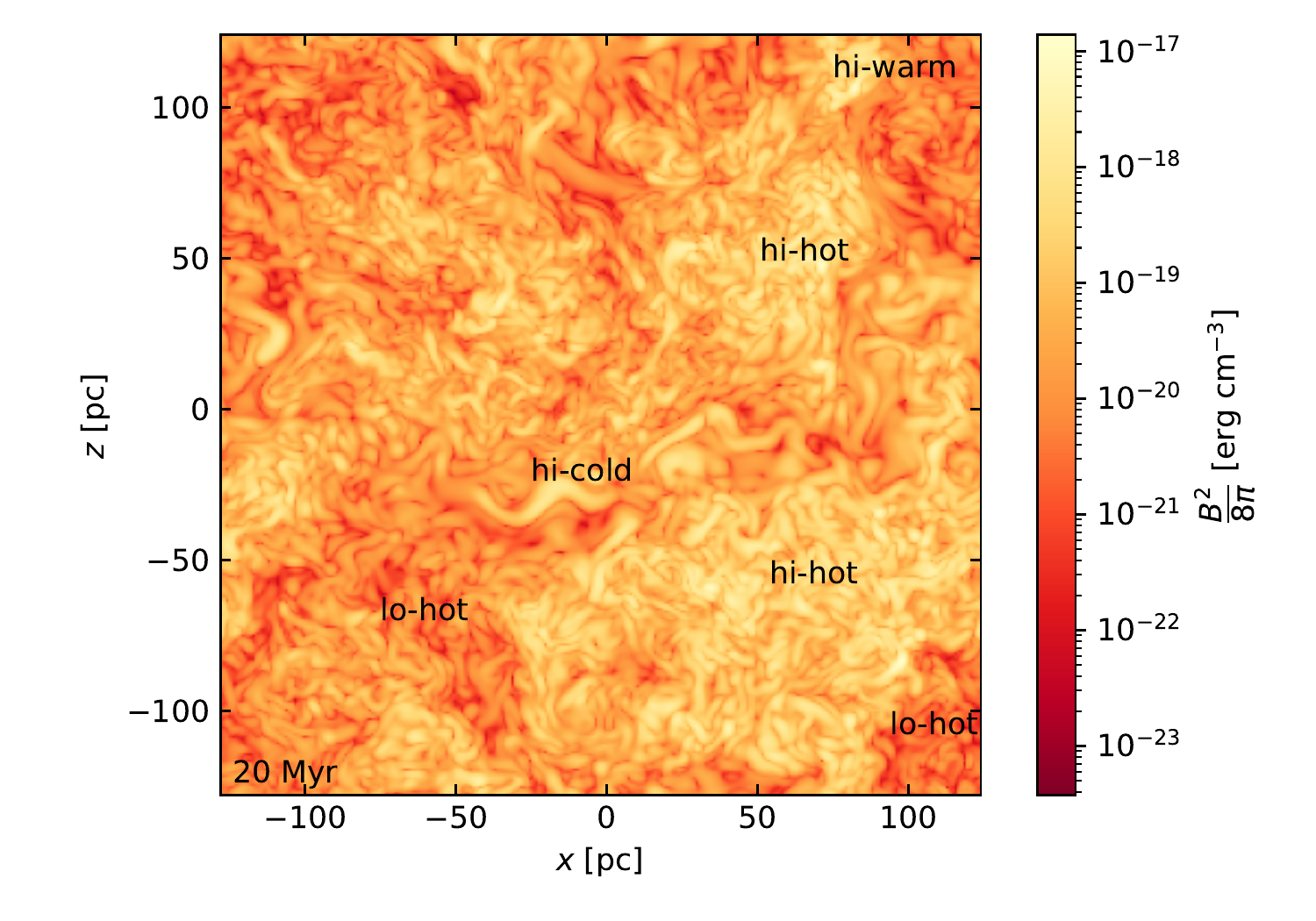}
  \begin{picture}(0,0)(0,0)
    \put(-265,590){{\sf\bf{(a)}}}
    \put(-265,395){{\sf\bf{(b)}}}
    \put(-265,200){{\sf\bf{(c)}}}
  \end{picture}
\caption{
	Slices from Model \hIIl\ of {\em (a)} gas number density, {\em (b)}
	temperature and {\em (c)} magnetic energy density, during an epoch of
	rapid magnetic growth.  At most times magnetic energy is concentrated
	in cold and warm phases, somewhat correlated with gas density,
	annotated as `hi-cold', `hi-warm' and `lo-hot'.  During this period of
	fast growth, however, magnetic energy is high in hot diffuse gas
	annotated by `hi-hot'. 
\label{fig:slices}
}
\end{figure*}
\subsection{Sorting time by growth rate}
\label{subsec:growth-epochs}

To examine how properties of the ISM in each phase differ between times with
rapid growth and slow growth or decay, for each model we compute the average
growth rate $\gamma(t)$ as the rate of change in $\eB$ during the specified
intervals from Table~\ref{tab:sampling}.  The intervals are chosen at each
resolution to exclude initial transient magnetic energy decay, prior to the
onset of any SSD, and most of the subsequent saturated dynamo stage, so that we
capture data of most relevance to SSD. The time series is then binned according
to where $\gamma(t)$ is lowest, median and highest.\footnote{ Sampling only the
lowest, median, and highest 5\%, 15\%, or 25\% yielded results consistent with
our method of simply splitting the growth rates into three bins.}

Histograms for each phase from snapshots at times belonging to each time bin
are then accumulated.  Fractional volume of SSD activity $\overline\fV$ for
each phase $i$ for a cumulative histogram is the mean of $\fV$ in snapshots
contributing to that histogram. Summary statistics of mean $\log \Gamma$ or
$\log \tilde\Gamma$ and relevant averaged physical quantities are calculated
from cumulative histograms in each bin.

\section{Results} \label{sec:results}

A hint toward explanation for the intermittent growth in the SSD in these
models appears on inspection of slices of the simulation data as displayed in
Figure~\ref{fig:slices} for Model \hIIl.  The expected response of the magnetic
energy to compressive flows is evident in regions highlighted by labels
`hi-cold', `hi-warm' and `lo-hot'.  What is anomalous in this scenario are the
regions highlighted as `hi-hot'.  In these regions the strongest magnetic field
is associated with the most diffuse and hottest ISM.  This cannot be explained
by passive compression of the field and suggests strong SSD activity in these
regions.  The snapshot shown is at $20\Myr$, a period in the simulation when
there is a burst of magnetic field growth.  Why is SSD present in these and not
in other regions of hot gas, and why is field growth strong during this period
and not others?

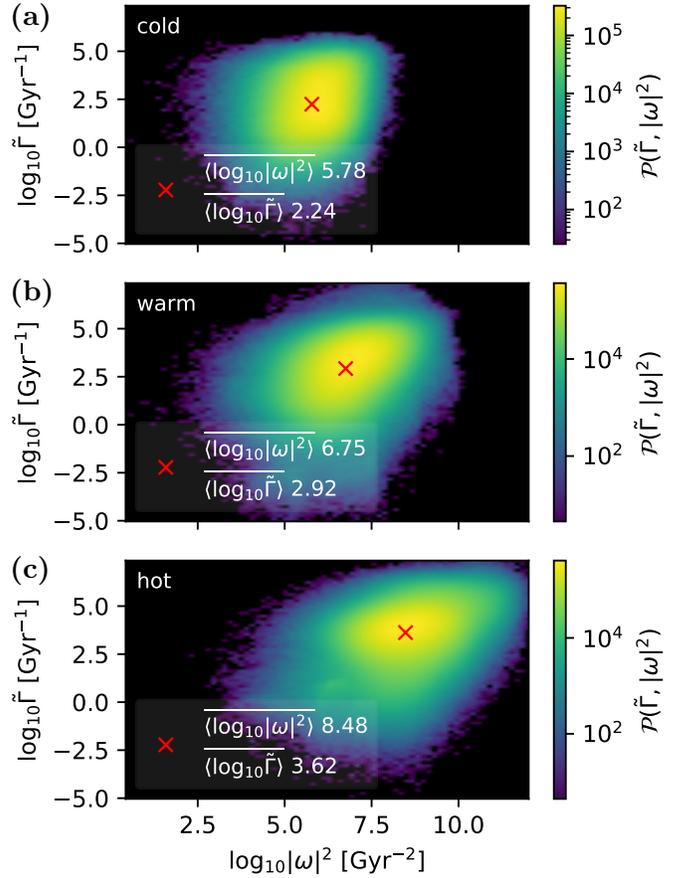
\begin{figure}
\centering
\includegraphics[trim=0.1cm 0.5cm 0.4cm 0.35cm,clip=true,width=0.48\textwidth]{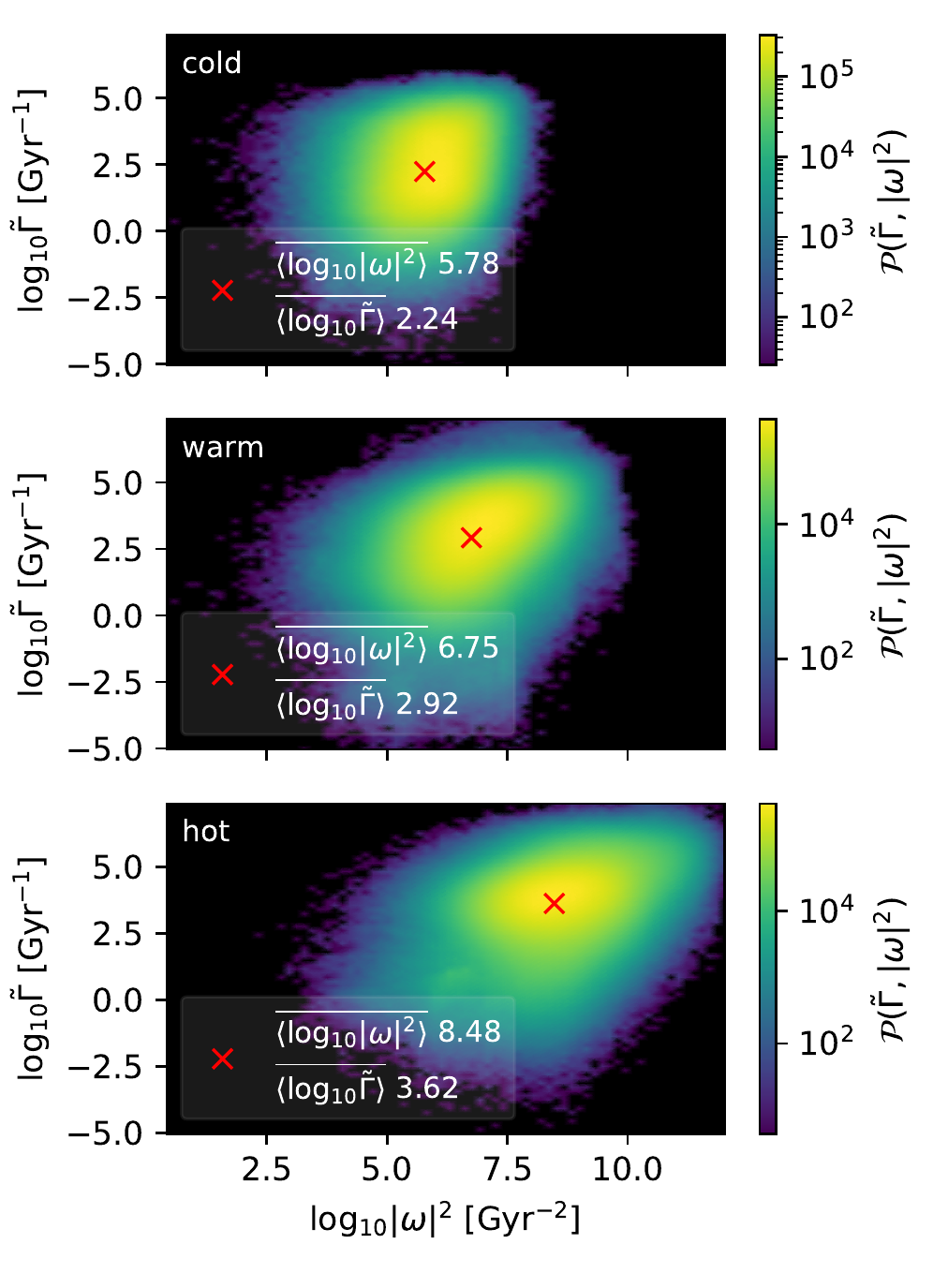}
  \begin{picture}(0,0)(0,0)
    \put(-120,336){{\sf\bf{(a)}}}
    \put(-120,232){{\sf\bf{(b)}}}
    \put(-120,127){{\sf\bf{(c)}}}
  \end{picture}
\caption{
	Histograms of log absolute magnetic energy growth rates $\tilde\Gamma$
	(Eq.~[\ref{eq:tildeG}]) vs log norm squared vorticity $|\omega|^2$ for
	Model \hVl\ for {\em (a)} cold gas, {\em (b)} warm gas and {\em (c)}
	hot gas.  The orange cross identifies the mean $\log \tilde\Gamma$ and
	$\log|\omega|^2$ of each distribution.  The histograms are cumulative
	results for all snapshots from the time intervals listed in
        Table~\ref{tab:sampling}, in this case $15<t<55$~Myr. \label{fig:histograms}
}
\end{figure}

For our core analysis of the multiphase structure of the SSD we primarily focus
on our models with resolution of 1~pc.  We include the highest resolution
0.5~pc model to demonstrate how well our solutions converge. The lower
resolution runs support insights into the dependence of the SSD on ISM
structure and SN rate.

We hypothesise that the erratic behaviour of the SSD is due to the changing
multiphase structure of the ISM.  To test this hypothesis, we compute joint
histograms by thermal phase of various physical properties in the total domain
from snapshots of each model alongside growth rates computed using
Equations~\eqref{eq:logeB} and~\eqref{eq:tildeG}.

\begin{figure}
\includegraphics[trim=0.35cm 0.6cm 0.3cm 0.3cm,clip=true,width=0.5\textwidth]{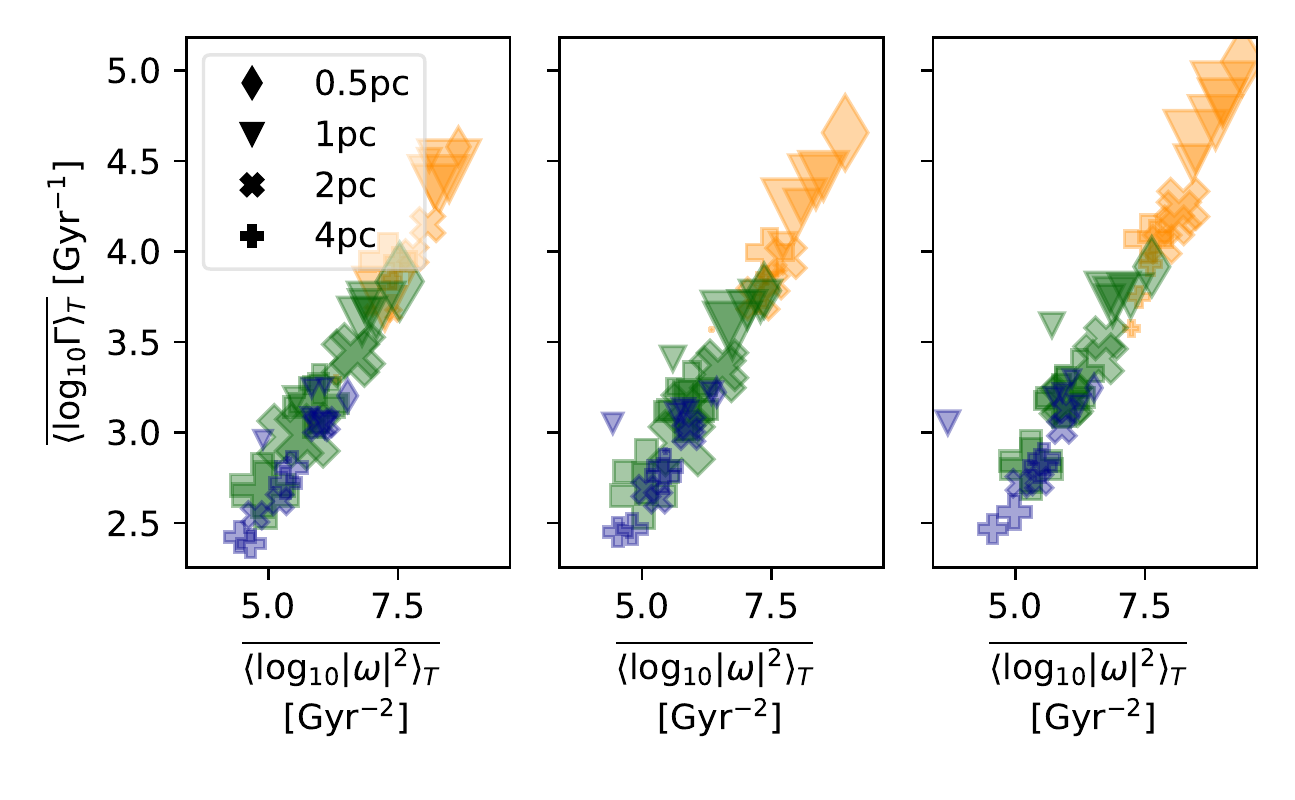}
\includegraphics[trim=0.15cm 0.6cm 0.3cm 0.3cm,clip=true,width=0.5\textwidth]{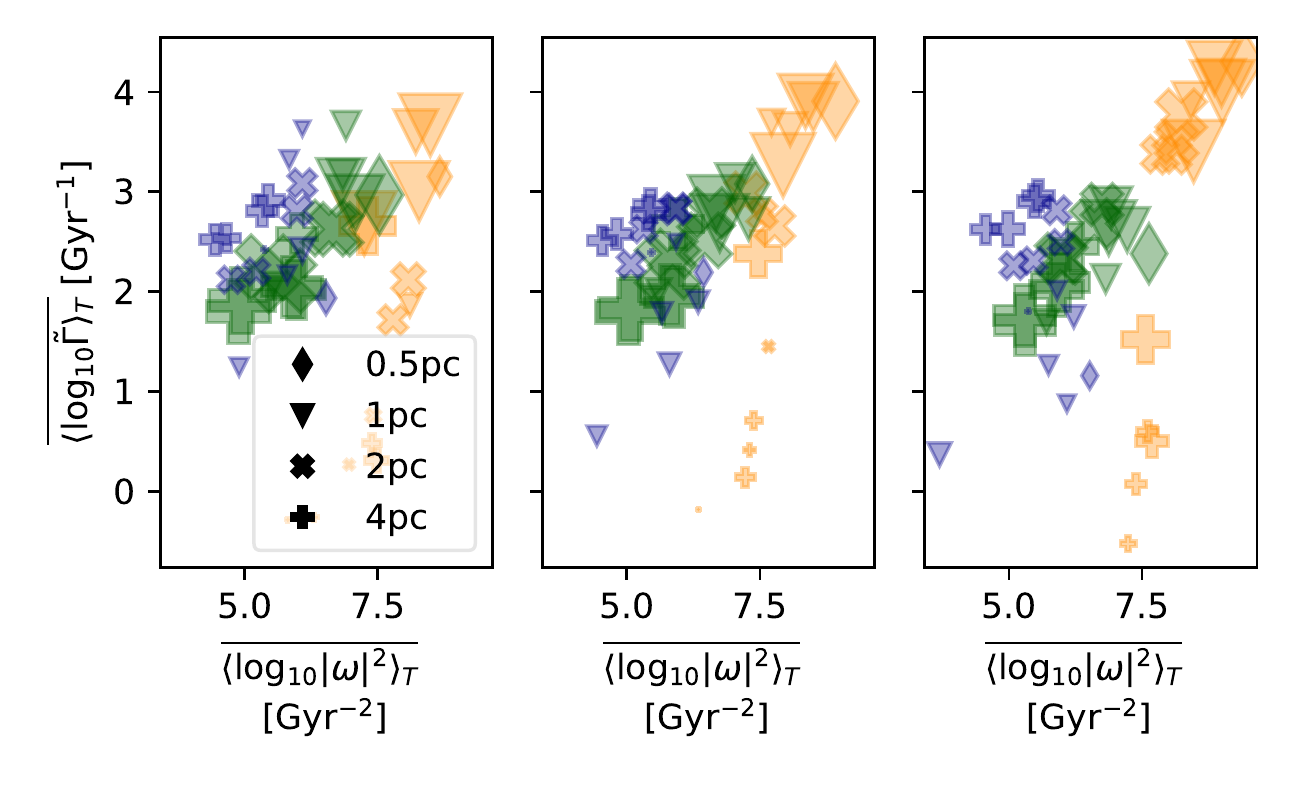}
  \begin{picture}(0,0)(0,0)
    \put(-5,305){{\sf\bf{(a)}}}
    \put(-5,150){{\sf\bf{(b)}}}
  \end{picture}
  \caption{
	Summary statistics of mean log norm squared vorticity $|\omega|^2$ for
all runs compared to mean log magnetic energy growth rates {\em (a)} relative
$\Gamma$ (Eq.~[\ref{eq:logeB}]) and {\em (b)} absolute $\tilde\Gamma$
(Eq.~[\ref{eq:tildeG}]).  Averages use the cumulative histograms for each phase
(denoted by angle brackets with T subscript; see Tab.~\ref{tab:sampling}).
Histograms accumulate snapshots binned by the \emph{(left)} lower,
\emph{(center)} median and \emph{(right)} upper growth rate $\gamma(t)$, as
explained in Sect.~\ref{subsec:growth-epochs}. The phase is indicated by
color: cold \emph{(blue)}, warm \emph{(green)}, and hot \emph{(orange)}. The
resolution is identified by shape as shown in the legend.  The fractional
volume $\overline\fV$ is proportional to the symbol area.
\label{fig:summary-vort}
}
\end{figure}
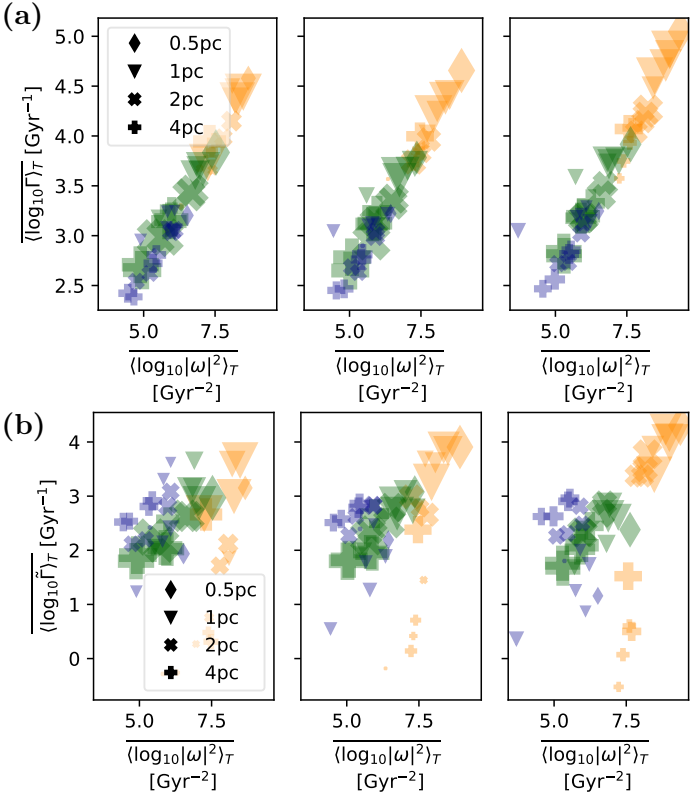

 \subsection{Vorticity}

In Figure~\ref{fig:histograms} we display the set of cumulative joint
histograms of absolute growth $\tilde\Gamma$ and (twice the) enstrophy or the
norm squared vorticity $|\omega|^2$ of cold, warm and hot gas from Model \hVl.
As is common across all models and variables, there is large variance and
histograms overlap between the phases.  However, a clear trend appears of
increasing vorticity and growth rate  with temperature. In the warm and hot
phases there is discernible positive correlation between vorticity and growth
rate within the phase. The phases and time intervals used for the histograms
are listed in Table~\ref{tab:sampling}.

The most striking correlation we find is between relative growth rate $\Gamma$
and the norm squared vorticity $|\omega|^2$, which we show in
Figure~\ref{fig:summary-vort}(a).  We find that the growth rate $\log\Gamma$ is
strongly proportional to $\log|\omega|^2$, and vorticity increases as we move
from cold to hot gas at all times.  At times belonging to the high growth rate
$\gamma(t)$ bin (right panel) the hot gas increases in vorticity, while the
vorticity in the cold and warm gas does not differ so much between bins.  The
fractional volume of the hot gas increases slightly for high resolution within
the upper bin.  So the efficiency of the SSD is linked to high vorticity and
high local and average growth rates $\Gamma$ and $\gamma$ are correlated with
increased vorticity in the hot gas\rev{, consistent with results from models using
more idealised turbulence \citep{FCSBKS11,AFTB21}}

To confirm whether this correlation is reflected in the absolute gains in
magnetic energy we plot the same summary for $\tilde\Gamma$ in
Figure~\ref{fig:summary-vort}(b).  At high resolution the same trends are
preserved as for $\Gamma$, with the changes in $\tilde\Gamma$ most associated
with changes in the characteristics of the hot gas. However for low resolution
$\tilde\Gamma$ is actually anti-correlated with vorticity. Increased mixing at
low resolution dampens vorticity and inhibits the formation of hot gas, damping
the SSD.

Energy growth at low resolution is better correlated with gas density, the cold
and then warm phases exhibiting higher $\tilde\Gamma$, consistent with results
from isothermal modelling, except at $\delta x=2\pc$ for the upper bin of
$\gamma(t)$ (right) when vorticity and fractional volume increases for the hot
gas.  Mean vorticity in the warm and cold phases does not change markedly
between bins.

\subsection{Mach number and Rm dependence}\label{sec:Rm}
\begin{figure}
\includegraphics[trim=0.35cm 0.6cm 0.3cm 0.3cm,clip=true,width=0.5\textwidth]{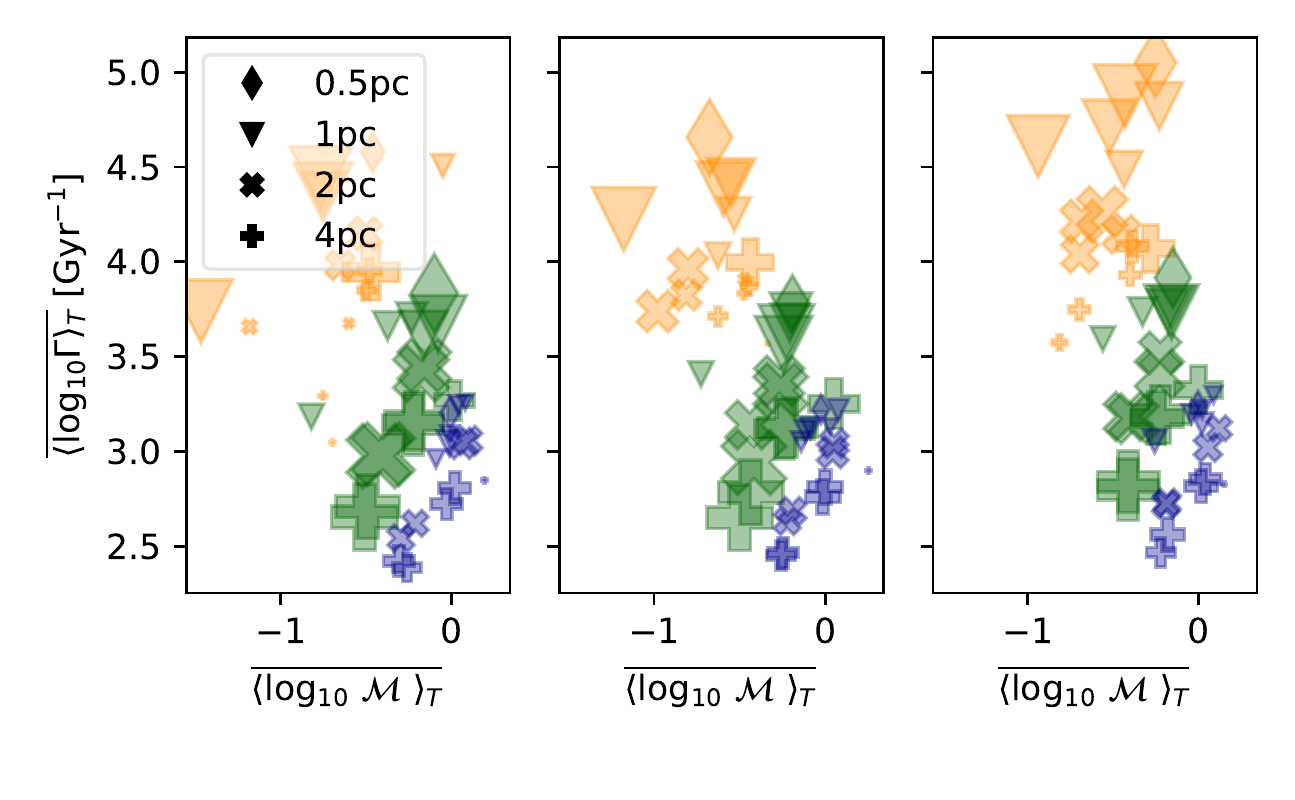}
\includegraphics[trim=0.35cm 0.6cm 0.3cm 0.3cm,clip=true,width=0.5\textwidth]{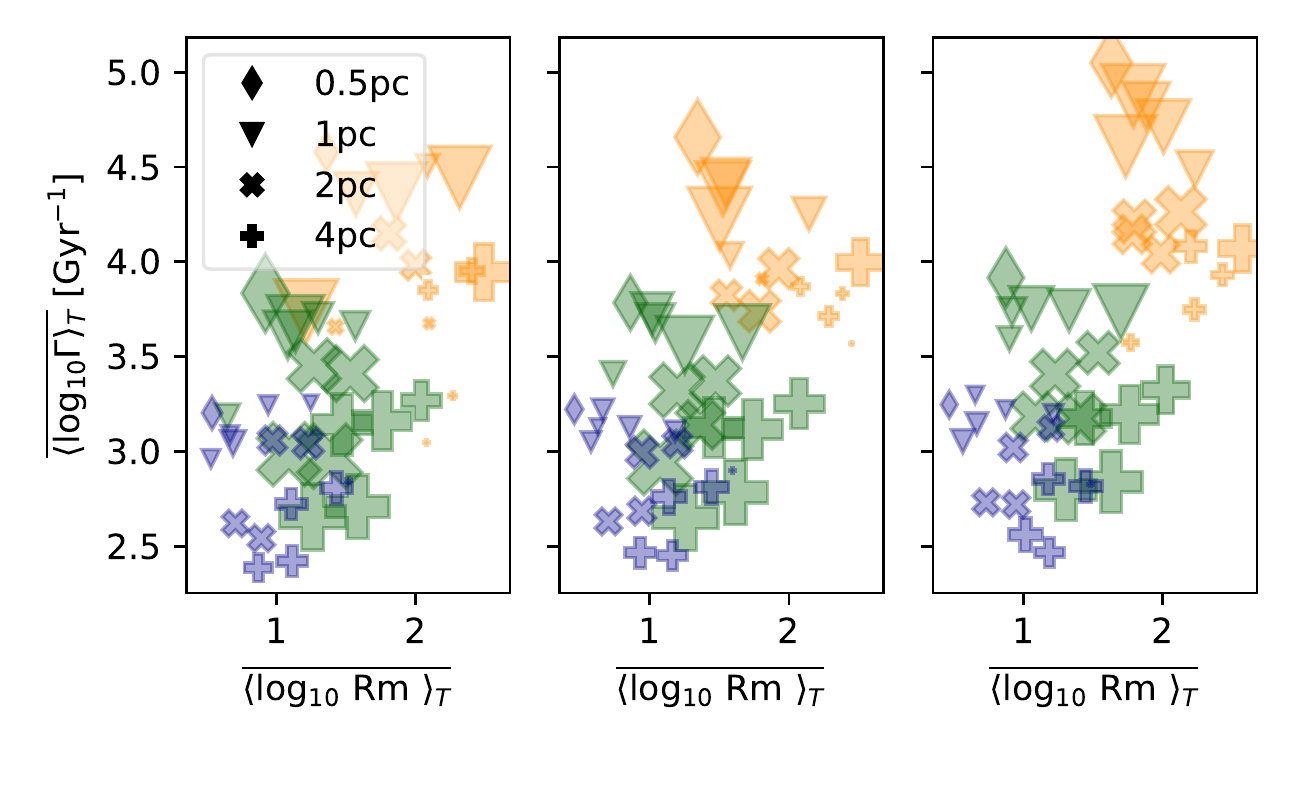}
\includegraphics[trim=0.35cm 0.6cm 0.3cm 0.3cm,clip=true,width=0.5\textwidth]{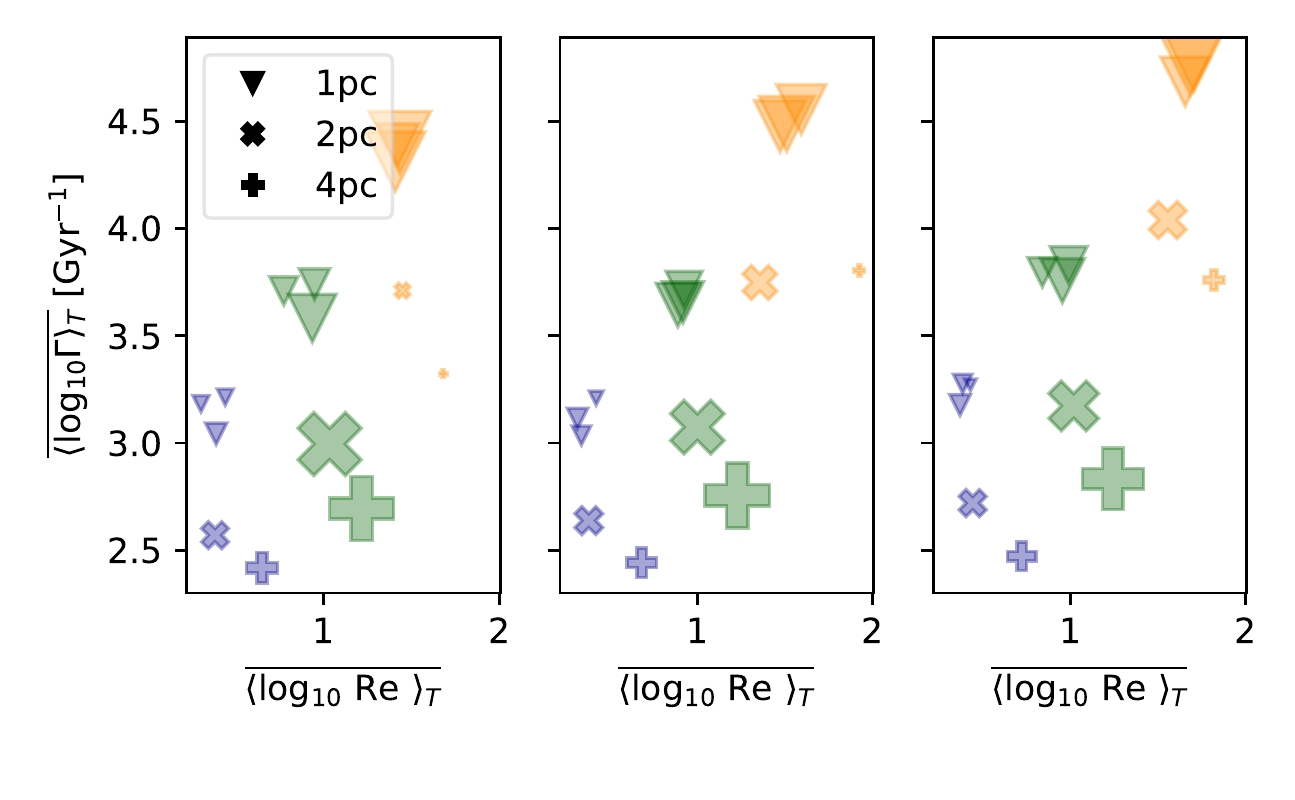}
  \begin{picture}(0,0)(0,0)
    \put(-5,460){{\sf\bf{(a)}}}
    \put(-5,305){{\sf\bf{(b)}}}
    \put(-5,150){{\sf\bf{(c)}}}
  \end{picture}
\caption{
	Summary statistics of the correlation of mean
	log $\Gamma$ to {\em (a)} mean log Mach number \Ms\rev{,} 
	{\em (b)} mean log magnetic Reynolds number $\Rm$ for all runs 
        \rev{and {\em (c)} mean log fluid Reynolds number}.
        The statistics are obtained, and symbols and panels have the same
        interpretation, as in Figure~\ref{fig:summary-vort}.
	\label{fig:summary-Ms}
}
\end{figure}
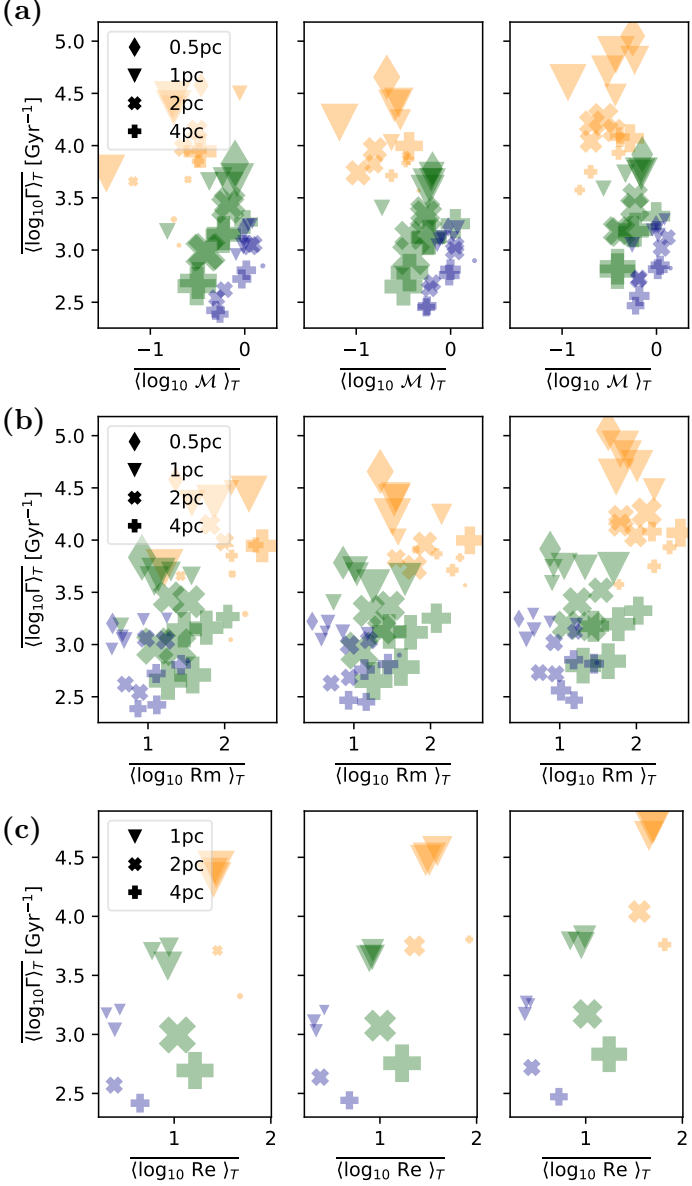

Results from models of isothermal SSDs indicate that highly compressible flows
inhibit dynamo activity \citep{Haugen:2004M,FCSBKS11,FSBS14,SF21}, while high
magnetic Reynolds number Rm makes SSDs more likely and is correlated with
higher growth rates \citep{SBK02,SHBCMM05,ISCM07,B11,Schober12,SBSW20,WKGR22}.

In Figure~\ref{fig:summary-Ms}(a) we display the summary statistics for Mach
number \Ms\ against relative growth rate $\Gamma$ for each bin of
volume-averaged magnetic energy growth rate.  Overall the expected trend of
$\Gamma$ reducing as \Ms\ increases is visible, more so in the lowest growth
rate regions (left panel).  We observe that hot gas with high sound speed has
low Ms and high relative growth rate $\Gamma$ and cold gas the inverse.  What
is, perhaps, unexpected is to see that within each phase there is a trend of
increasing $\Gamma$ with increasing \Ms.  This can be explained by the counter
effect of increased velocities, which drive higher Mach numbers for a given
sound speed, but also higher Rm.  So increased $\Gamma$ with \Ms\ should have
complementary trends in Rm.

\begin{figure}
\centering
\includegraphics[trim=0.1cm 0.4cm 0.4cm 0.35cm,clip=true,width=0.48\textwidth]{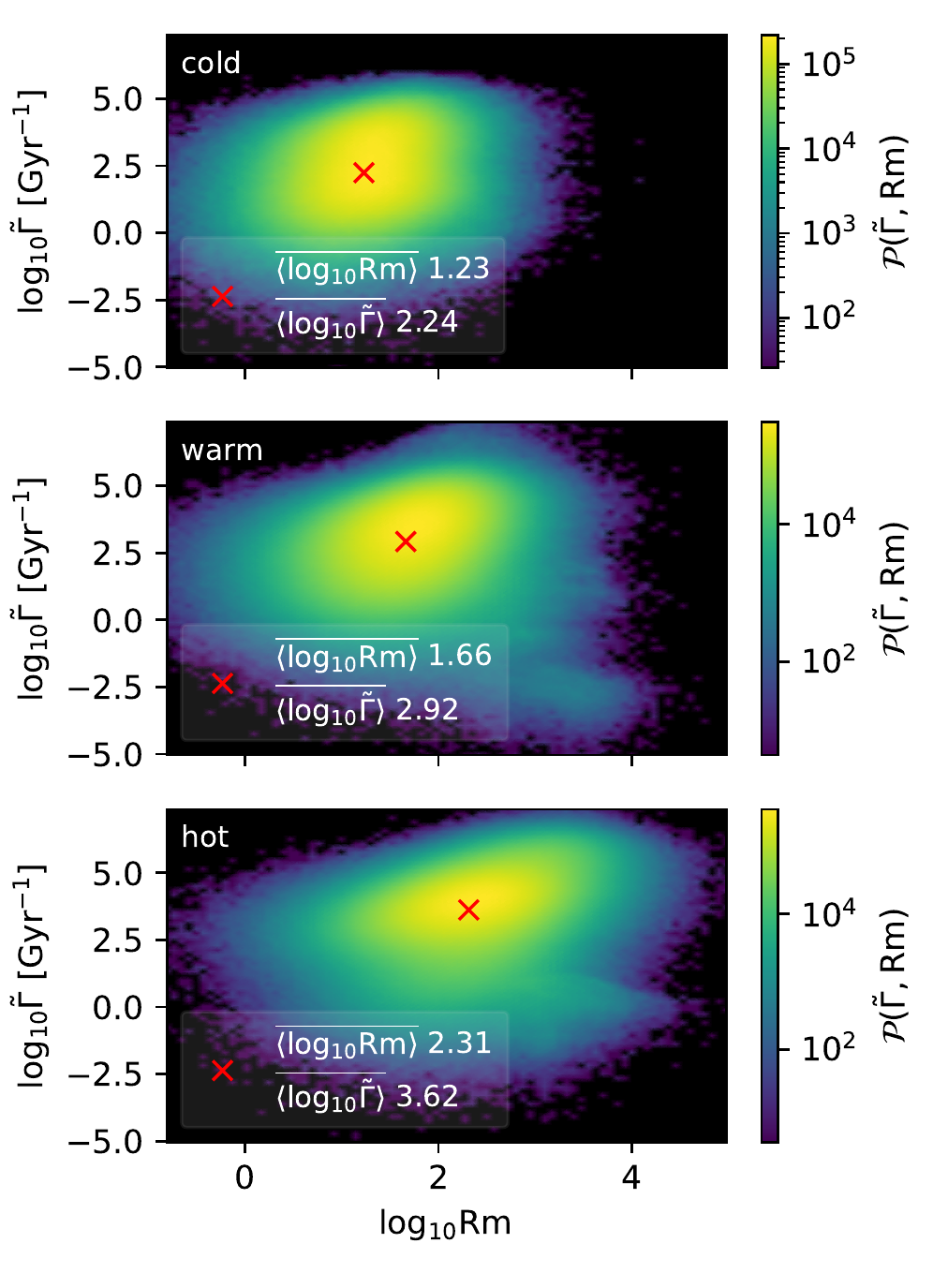}
  \begin{picture}(0,0)(0,0)
    \put(-120,339){{\sf\bf{(a)}}}
    \put(-120,233){{\sf\bf{(b)}}}
    \put(-120,127){{\sf\bf{(c)}}}
  \end{picture}
\caption{
        Histograms of log absolute growth rates
	$\tilde\Gamma$, Eq.~\eqref{eq:tildeG}, vs log Rm, Eq.~\eqref{eq:Rm},
	for Model \hVl. Otherwise as described in Figure~\ref{fig:histograms}.
	\label{fig:Rmhist}
}
\end{figure}
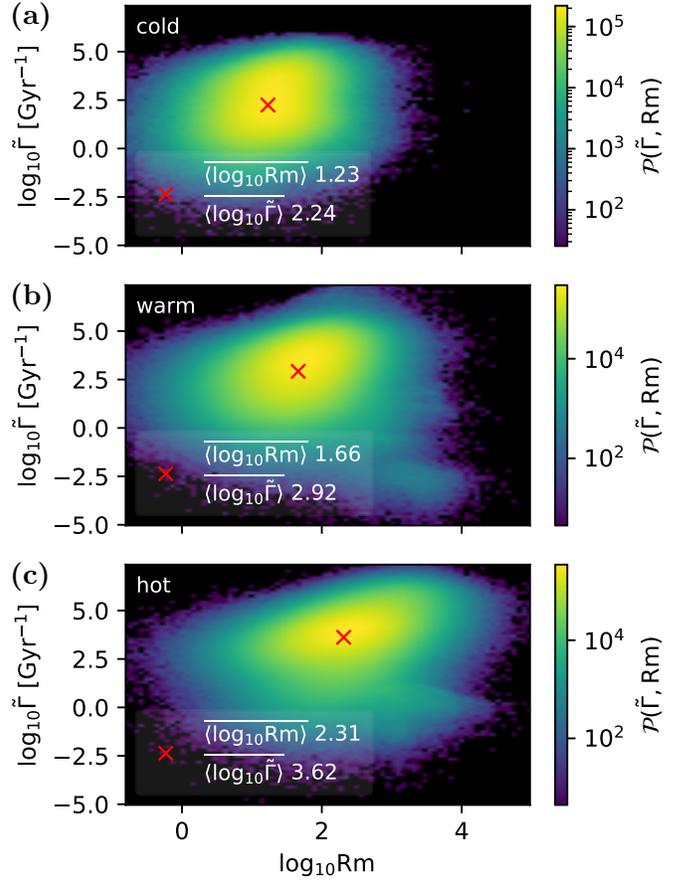

In this highly compressible system it is challenging to calculate Rm as would
conventionally be defined for weakly compressible simulations. Instead, we use
the ratio of advective to diffusive terms given by Eq.~(\ref{eq:Rm}).  With
strongly fluctuating characteristic velocities and length scales applying
between the continuously shifting phases, and the very sporadic and flexible
scale of the SN forcing, Rm varies as a function of position and time.  An
illustration of the extent to which Rm varies is shown in the histograms of
Figure~\ref{fig:Rmhist}.  While the mean Rm is quite modest of order 16 and 45
in the cold and warm phases, there are regions in all phases where Rm~$> 10^3$,
and in the hot phase Rm~$> 10^4$.  For the summary statistics the log mean
values are quite informative to identify the trends in SSD dependencies across
simulations over time and between phases.  

Results for Rm versus $\Gamma$ are shown in Figure~\ref{fig:summary-Ms}(b).
Velocities increase with temperature, increasing Rm and supporting faster SSD,
an overall trend as would be expected for $\Gamma$.  Within each phase and
within each $\gamma(t)$ bin, however, $\Gamma$ is weakly anticorrelated with
Rm, and across each bin the patterns are largely unaffected for cold and warm
gas, suggesting higher velocities may be associated with higher Mach numbers.
\rev{Results are similar for Re versus $\Gamma$ as shown in
Figure~\ref{fig:summary-Ms}(c) for samples from models that include explicit
viscosity $\nu$.  The higher resolution models ($\delta x = 1$) show a good
correlation between Re and growth rate.  The lower normalization of that
correlation seen in the lower resolution models may well be due to 
\rev{reduced kinetic energy from excess cooling in under-resolved density gradients.} 
The comparison between the 1~pc and 0.5~pc models without viscosity
suggests that the 1~pc model is well converged.} For the hot gas Rm\rev{, Re,}
and $\Gamma$ \rev{all} increase for the upper growth rate bin, suggesting
increased velocities are associated with rotational rather than compressive
flows.
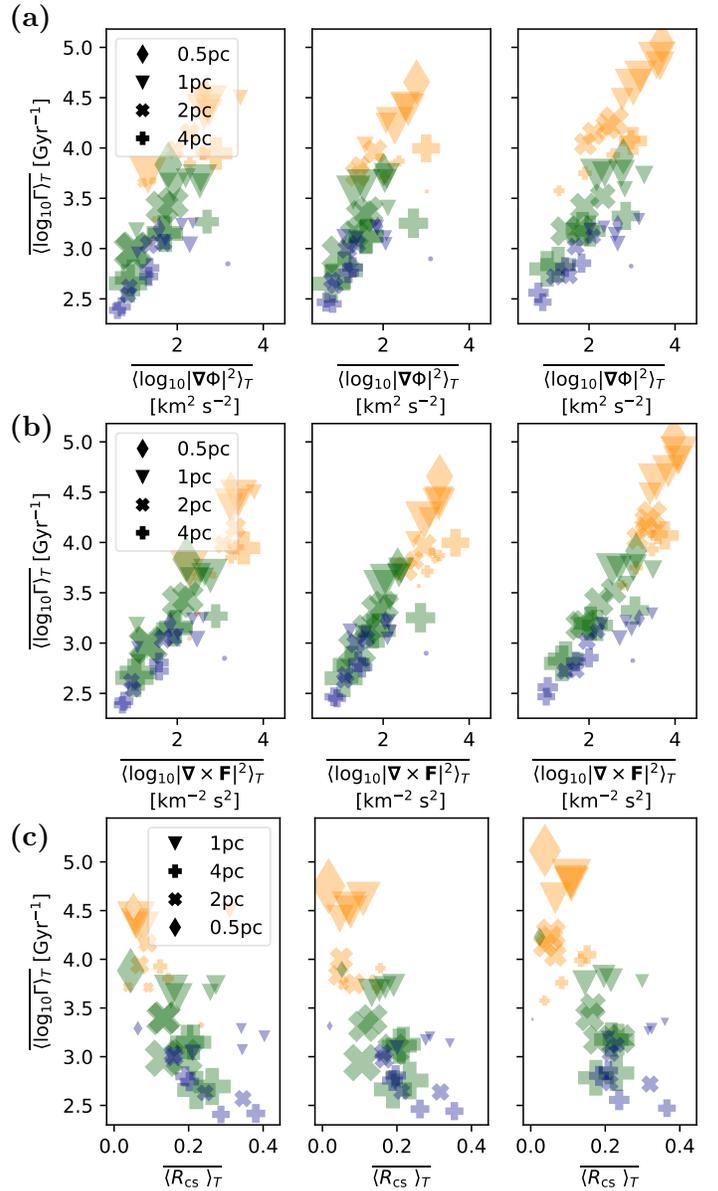
\begin{figure}
\includegraphics[trim=0.35cm 0.6cm 0.3cm 0.3cm,clip=true,width=0.5\textwidth]{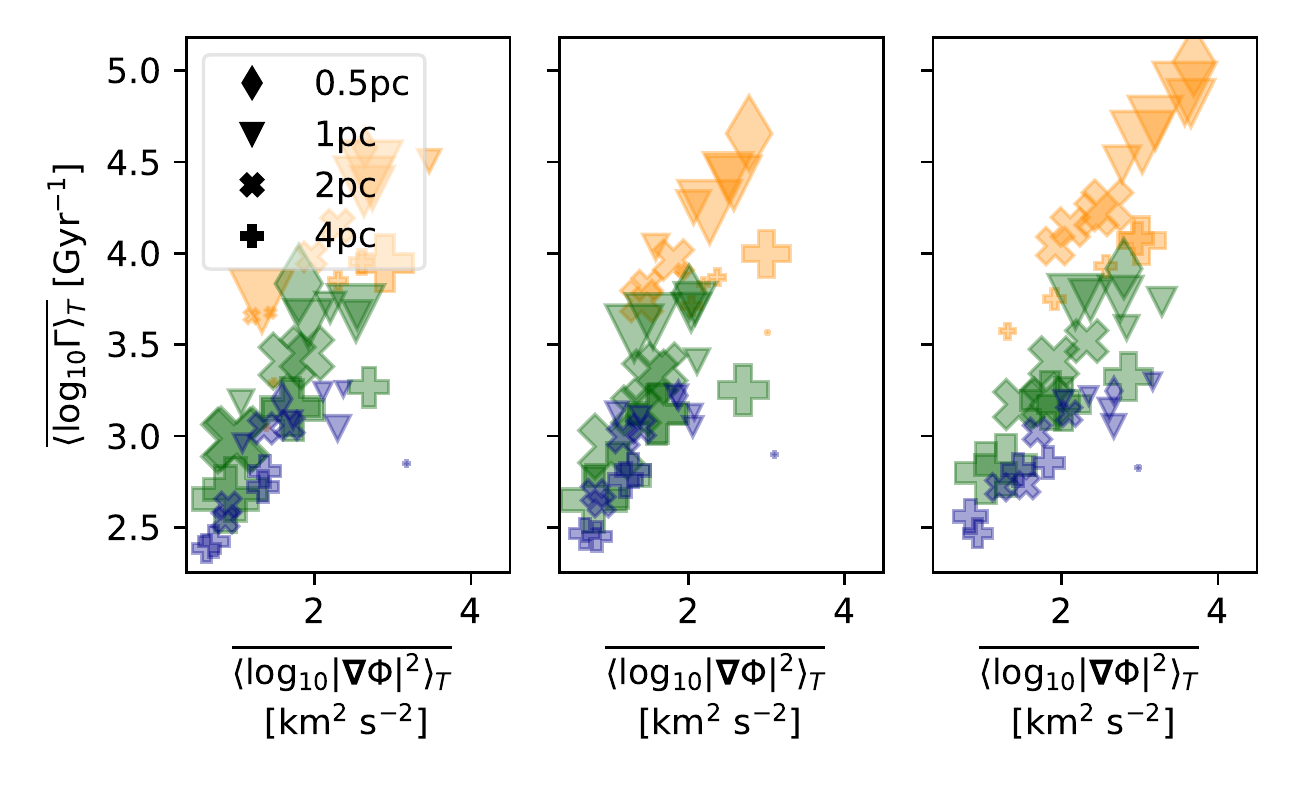}
\includegraphics[trim=0.35cm 0.6cm 0.3cm 0.3cm,clip=true,width=0.5\textwidth]{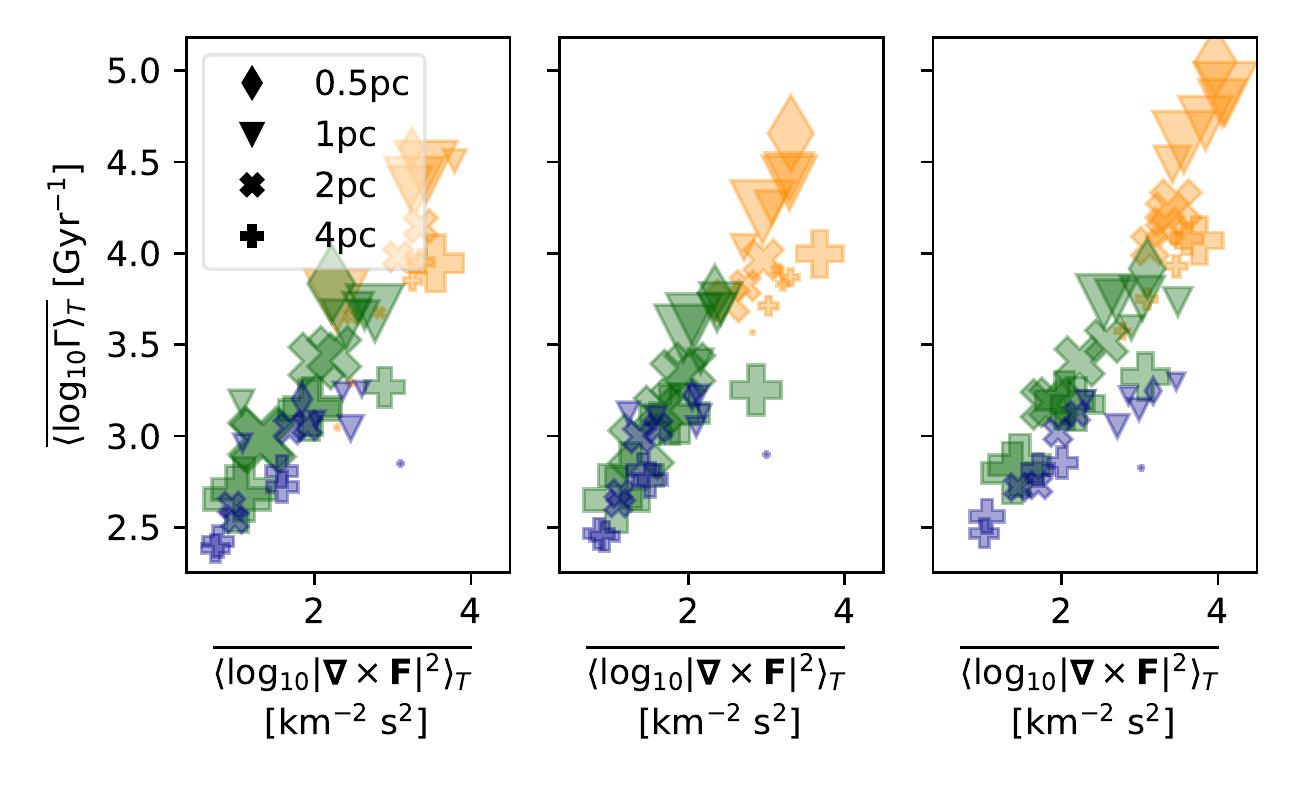}
\includegraphics[trim=0.35cm 0.6cm 0.3cm 0.3cm,clip=true,width=0.5\textwidth]{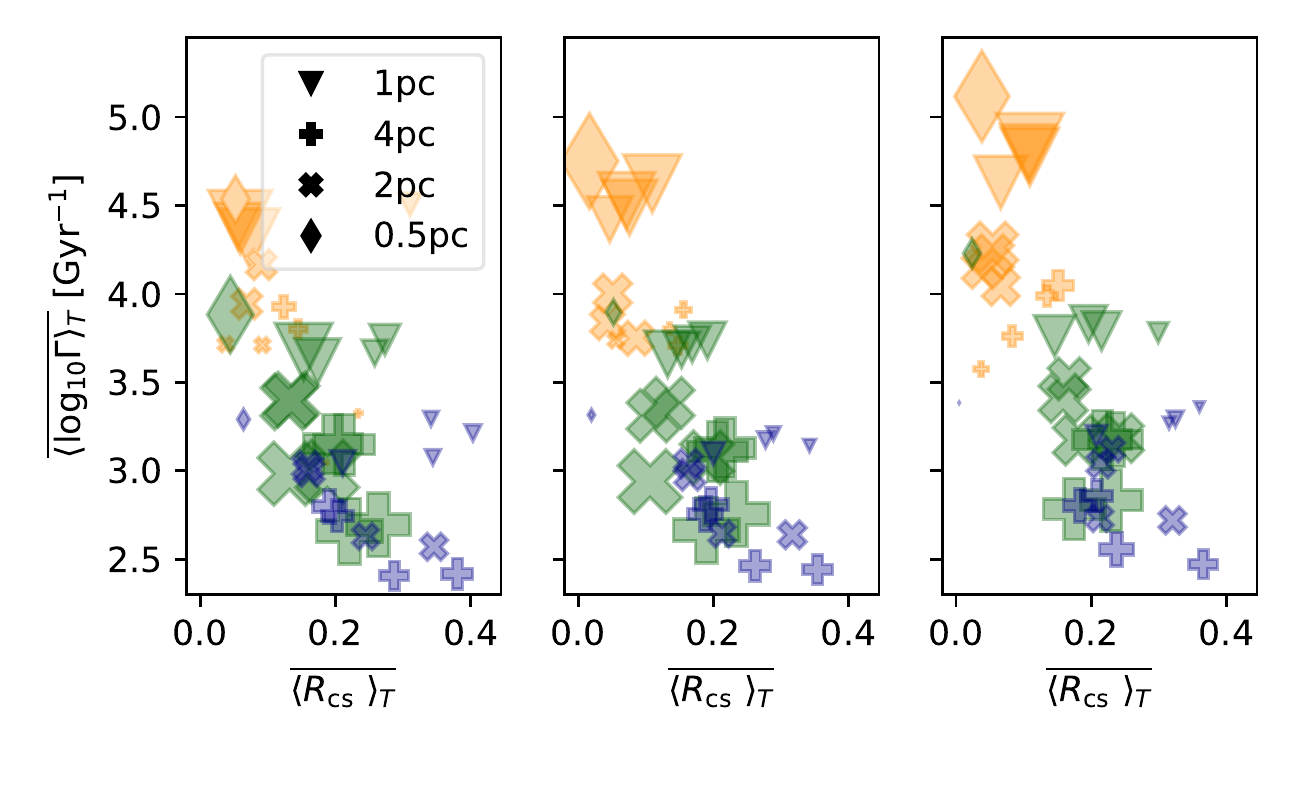}
  \begin{picture}(0,0)(0,0)
    \put(-5,460){{\sf\bf{(a)}}}
    \put(-5,305){{\sf\bf{(b)}}}
    \put(-5,150){{\sf\bf{(c)}}}
  \end{picture}
\caption{
	 Summary statistics of mean log relative
	 growth rates $\Gamma$ vs.\ mean log norm squared {\em (a)} compressive \rev{and
	 {\em (b)} rotational velocities 
	 and {\em (c)} mean compressive ratio $R_{\rm cs}$} for all runs. 
        The statistics are obtained, and symbols and panels have the same
        interpretation, as in Figure~\ref{fig:summary-vort}.
	 \label{fig:summary-urot}
}
\end{figure}

If this were the case, we would expect to see this reflected in the structure
of the flow, which we consider in Figure~\ref{fig:summary-urot} \citep[see
analysis of hydrodynamic flow by][]{KGVS18}.  We use a Helmholtz decomposition
to separate the velocity field $u=u_1+u_2$ into purely solenoidal
$u_1=\nabla\times\vect{F}$ and compressive flows $u_2=\nabla\Phi$, in which
$\vect{F}$ and $\Phi$ are, respectively, a vector potential and scalar
potential pair from which orthogonal flows $u_1$ and $u_2$ can be derived.

Both flows appear strongly correlated with $\Gamma$, although not as clearly as
vorticity in Figure~\ref{fig:summary-vort}(a).  It is reasonable to expect that
the magnitude of compressive and solenoidal flows would be correlated, and as
such some common correlation to $\Gamma$ would be evident.  The clearer
correlations for (b) rotational rather than (a) compressive flows indicates
that the latter is coincidental.  Growth rate trends in the warm and cold gas
do not alter significantly between bins of $\gamma(t)$ even as flow strength
varies.  In the hot gas $\Gamma$ clearly increases when
$|\nabla\times\vect{F}|$ increases between low (left) and high (right)
$\gamma(t)$ bins, especially for high resolution models.

\rev{In Figure~\ref{fig:summary-urot}(c) the summary statistics are plotted
of the compressive ratio \citep{KNPW07} defined by}
\begin{equation}\label{eq:cratio}
\rev{R_{\rm cs} =
	\frac{\langle|\nabla\cdot\vect{u}|^2\rangle}{
		\langle|\nabla\cdot\vect{u}|^2\rangle+\langle|\nabla\times\vect{u}|^2\rangle}.}
\end{equation}
\rev{This shows that the ratio of solenoidal energy to total compressive and
solenoidal energy $(1-R_{\rm cs})$ is typically between 60\% (cold) and 90\%
(hot).  At higher resolution the fraction of energy in \rev{solenoidal flow }
is higher. The compressive ratio reduces from cold to hot phases.  \rev{The
growth rate is directly correlated with the fraction of solenoidal energy.} }

\subsection{Prandtl number variability}

\begin{figure*}
\centering
\includegraphics[trim=0.4cm 0.5cm 0.0cm 0.0cm,clip=true,width=0.6\textwidth]{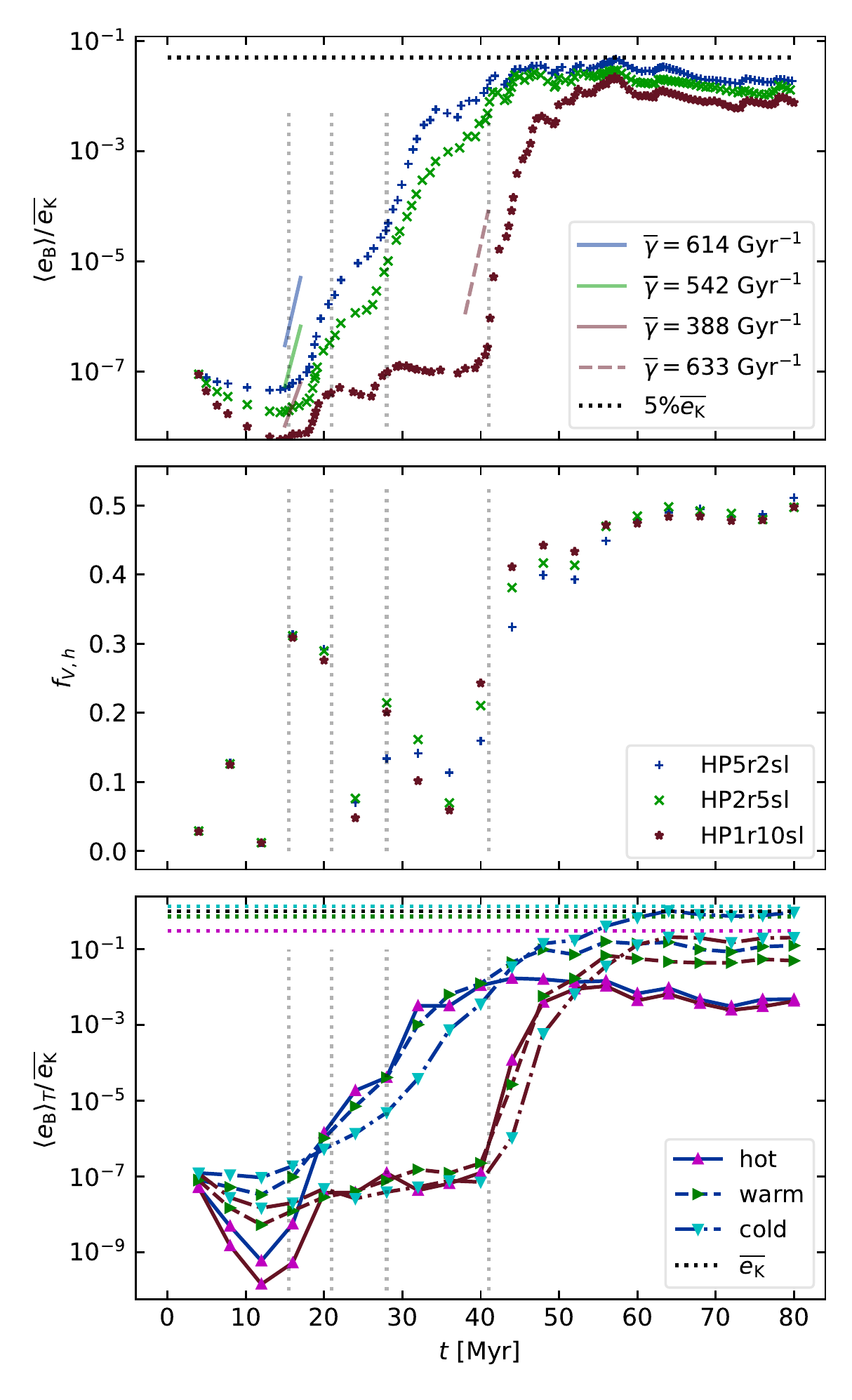}
  \begin{picture}(0,0)(0,0)
    \put(-320,480){{\sf\bf{(a)}}}
    \put(-320,320){{\sf\bf{(b)}}}
    \put(-320,160){{\sf\bf{(c)}}}
  \end{picture}
\caption{
	{\em (a)} Evolution of magnetic energy density $\eB$ for \hIl\ {\em
	(brown)}, \hIIl\ {\em (green)} and \hVl\ {\em (blue)}, normalized by
	$\eK$, the hydrodynamic statistically steady average kinetic energy
	density at $8<t<32\Myr$, prior to saturation of the SSD.  Fits of
	growth rate $\barg$ for each model span a period over which growth can
	robustly be considered exponential: $18 < t<20$~Myr. A second fit for
	\hIl\ spans $41<t<44$~Myr.  The black horizontal dotted line at 5\% of
	$\eK$ indicates the energy at which SSD typically saturates in these
	models.  {\em (b)} The concurrent evolution of the fractional volume
	$\fH$ of hot gas.  {\em (c)} Evolution of $\eB_T$ by phase $T$ for
	\hIl\ and \hVl.  Line color is as {\em (b)}, and line styles and
	symbols denote thermal phase, as indicated in the legend. The values of
	$\eK_T/\eK$ for each phase are indicated by colored horizontal dotted
	lines, and the black horizontal line here indicates 100\% of $\eK$.
	Vertical dotted lines are included to identify significant epochs,
	which are subintervals of those intervals listed in
	Table~\ref{tab:sampling}. 
\label{fig:Pm}
}
\end{figure*}

The multiphase structure and intermittent forcing of the ISM causes the fluid
Reynolds number Re to vary just as was shown for Rm in Figure~\ref{fig:Rmhist}.
This results in a wide variation of magnetic Prandtl number Pm.  In
\citet[][see Figures 5{[b1]} and {[b2]}]{GMKS21} we examined the role of Pm on
the SSD in ISM simulations, by varying both viscosity $\nu$ and resistivity
$\eta$.  For a fixed $\eta$, the SSD did not depend on Pm and the energy at
which the dynamo saturated was consistently around 5\% of equipartition.  For a
fixed $\nu$, on the other hand, SSD growth rate did increase with Pm for
$\Pm\leq5$ in the parameter space considered.  The energy at saturation
increased with Pm, remaining steady for $\Pm\gtrsim 5$.  This suggested that it
is the ratio of the Lorentz force to Rm rather than Pm that is the controlling
parameter \citep[see also][for a similar suggestion in magnetorotational
dynamos]{oishi2011}.

Therefore, here we vary Pm by varying only $\eta$. Where the viscosity, $\nu$,
and the resistivity, $\eta$, are explicitly defined, we can nominally express
the magnetic Prandtl number, $\Pm=\nu/\eta$, which is what we shall use here.
Alternatively, as a function of position Pm can also vary due to the flows and
length scales characteristic at different regions and times altering the
effective ratios of Rm to Re.  In the Appendix we determine the effective
magnetic Prandtl number $\Pm_{\rm eff}$ in various models to examine the
relevance the explicit diffusivities to the dynamics (see
Figure~\ref{fig:power}).

In Figure~\ref{fig:Pm}(a) we show the volume averaged magnetic energy density
$\eB$ for models with $\Pm\in[1,5]$ with resolution $\delta x=1\pc$.  Around
$20\Myr$ there is a burst of SSD, common to all three runs, highlighted between
the two leftmost vertical dotted lines.  Growth rates $\barg$ applying for
$18<t<20\Myr$ depend on Pm in this range.  Sensitivity to Pm reduces as Pm
approaches 5, consistent with the idea that growth rates become asymptotic for
higher Pm.  We also fit $\barg$ for $\Pm=1$ at $41<t<44\Myr$, where the growth
rate exceeds that for $\Pm=5$ earlier.  Under most circumstances we would
expect $\barg$ to be lower for the larger $\eta$, but during this period the
fractional volume of the hot gas $\fV$ is higher (Figure~\ref{fig:Pm}(b)) in all
models than during the earlier subinterval.  A surge aligned to the rightmost
vertical dotted line is also evident in the high Pm models, although it is
inhibited in those by the onset of saturation.  

The critical Rm above which SSD can be excited and the typical scaling
relations with Rm for the rate of magnetic energy growth are not well
determined.  For $\Pm=1$ SSD actually decays during two periods, before its
final surge. This could be due to Rm varying or changing conditions resulting
in a lower critical Rm.  

We plot the hot gas fractional volumes $\fH$ for the models with $\Pm\in[1,5]$
in Figure~\ref{fig:Pm}(b).  At first, the hot gas fractions $\fH$ for all
models coincide.  Despite the initial weakness of magnetic effects, the
fractional volumes diverge slightly within 20~megayears. The resistive timestep
differs between the models; small changes in timestep lead to the chaotic
solutions adopting alternate trajectories.  For example a delay of a few
decades in scheduling a single SN, even at the identical location, alters the
specific ambient conditions modifying the remnant evolution and subsequent
dynamics, propagating into diverging trajectories between models. 

Nevertheless the trends are consistent between models.  In the subinterval
between the two leftmost vertical dotted lines, $\fH$ is enhanced,
corresponding to a burst in SSD activity.  The dip in $\fH$ between the middle
two vertical lines corresponds to decay in the $\Pm=1$ model and slower growth
at higher Pm, followed by another SSD burst aligned with the third vertical
line and a peak in $\fH$, least so for $\Pm=5$.  The value of $\fH$ then rises
again at 40~megayears, particularly for $\Pm=1$, with an accompanying boost in
SSD.  The described behavior supports the hypothesis that higher SSD activity
is correlated with increasing fractional volume of the hot gas $\fH$.

\subsection{Growth trends by phase}
To examine our hypothesis further, we split the evolution of the magnetic
energy between phases $\eB_T$ and plot these in Figure~\ref{fig:Pm}(c) for the
models with $\Pm=1$ and 5.  These are normalised by $\eK$.  Horizontal dotted
lines of matching color for each phase indicate the time-averaged kinetic
energy density by phase $\eK_T/\eK$.  These lines, plotted for $\Pm=5$, are
very similar for $\Pm=1$.

Usually $\eB_T$ is smallest in the hot gas, but during the SSD activity bursts,
its growth is the most rapid and its energy density then exceeds that of the
other two phases.  The saturation energy of the hot gas is similar for both
models, around $0.05 \eK_h$ at 40~megayears for $\Pm=5$, and 55~megayears for
$\Pm=1$.  In contrast the saturation energy in the cold and warm phases is
affected by $\Pm$.  In the nonlinear phase the magnetic energy in the hot phase
$\eB_h$ decays to around $0.01 \eK_h$, while the magnetic energy in the cold
phase $\eB_c$ grows, above $\eK_c$ in the case of $\Pm=5$.

\begin{figure}
\centering
\includegraphics[trim=0.3cm 0.5cm 0.0cm 0.0cm,clip=true,width=0.48\textwidth]{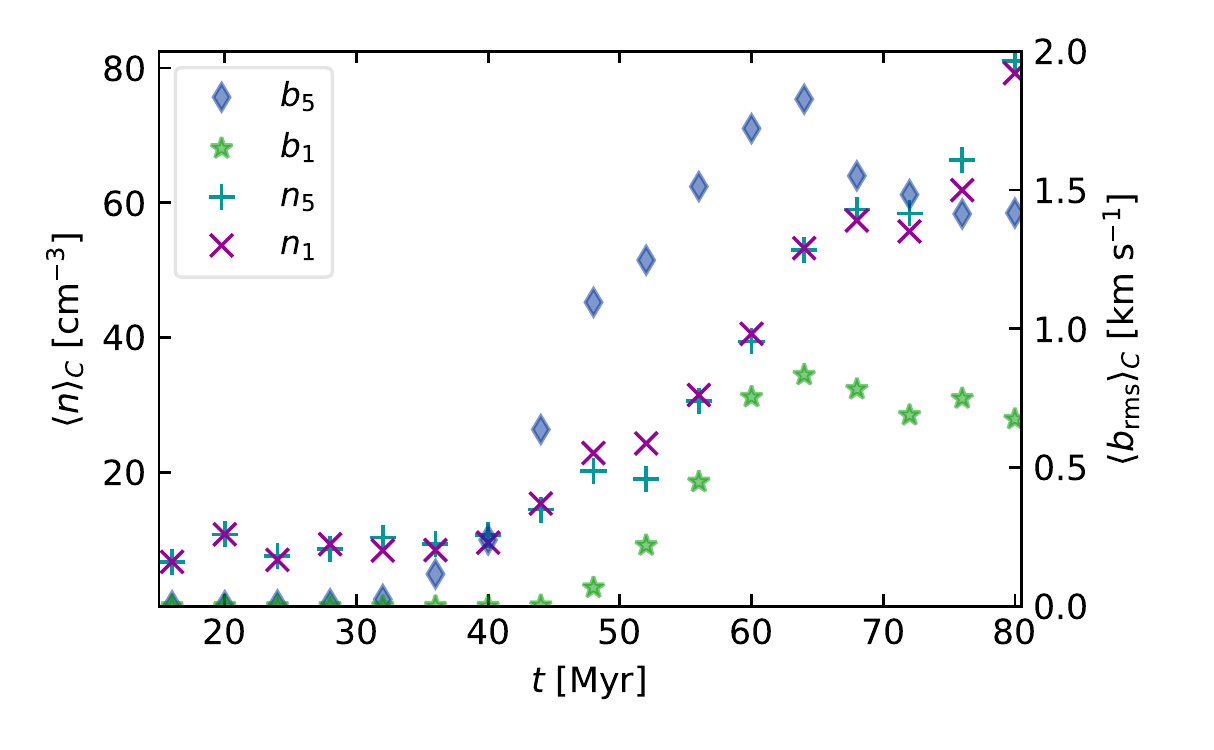}
\caption{
	\rev{For the cold gas in models of nominal $\Pm=1$ and 5, as displayed
	in Figure~\ref{fig:Pm}, the mean gas number density $\langle
	n\rangle_C$ {\em (left axis)} and root mean square mean magnetic field
	strength $\langle b_{\rm rms} \rangle_C$ {\em (right axis)} normalised
	by the square root of mean cold gas density. The subscripts indicate
	the nominal Pm.}
	\label{fig:tilden}
}
\end{figure}

Is the latter consistent with compression in the statistical steady MHD state?
Under compression, magnetic field strength relates to density $|B|\propto
\rho^\alpha$, where $\alpha \in[0,1]$ for compression along field lines up to
disc or slab-like compression \citep{TPMTP15}.  In Figure~\ref{fig:tilden}, for
the two models in Figure~\ref{fig:Pm}(c) we plot the \rev{mean gas number
density in the cold gas (left axis) and the root mean square strength of the
magnetic field normalised by cold gas density $\langle b_{\rm rms}\rangle_C
=\langle B^2\rangle_C^{1/2}\langle \rho\rangle^{-\alpha}_C$ (right axis),
taking $\alpha=1/2$.  In the interval $50\Myr<t<60\Myr$ the dynamo has clearly
saturated for the warm and hot gas (Figure~\ref{fig:Pm}).  If the continued
growth of the magnetic field in the cold gas were due to compression of the
field with the gas, then we would expect the ratio of $b$ to remain steady or
even decay.  Instead, in both models this ratio continues to increase until
$\sim65\Myr$, indicating that dynamo remains active in the cold phase.  The
increasing mean density in the cold gas toward the end of the simulation
reflects the increasing fractional volume of hot gas and mean thermal energy
density. The cooling in the hot gas is too inefficient to balance the supply of
thermal energy from subsequent SNe in the closed system, and thermal runaway
will eventually result.  Applying $2/3<\alpha<1$ in the analysis does not
qualitatively alter the development of $b$.}

\subsection{Supernova rate}

\begin{figure}
\centering
\includegraphics[trim=0.0cm 0.0cm 0.0cm 0.0cm,clip=true,width=0.49\textwidth]{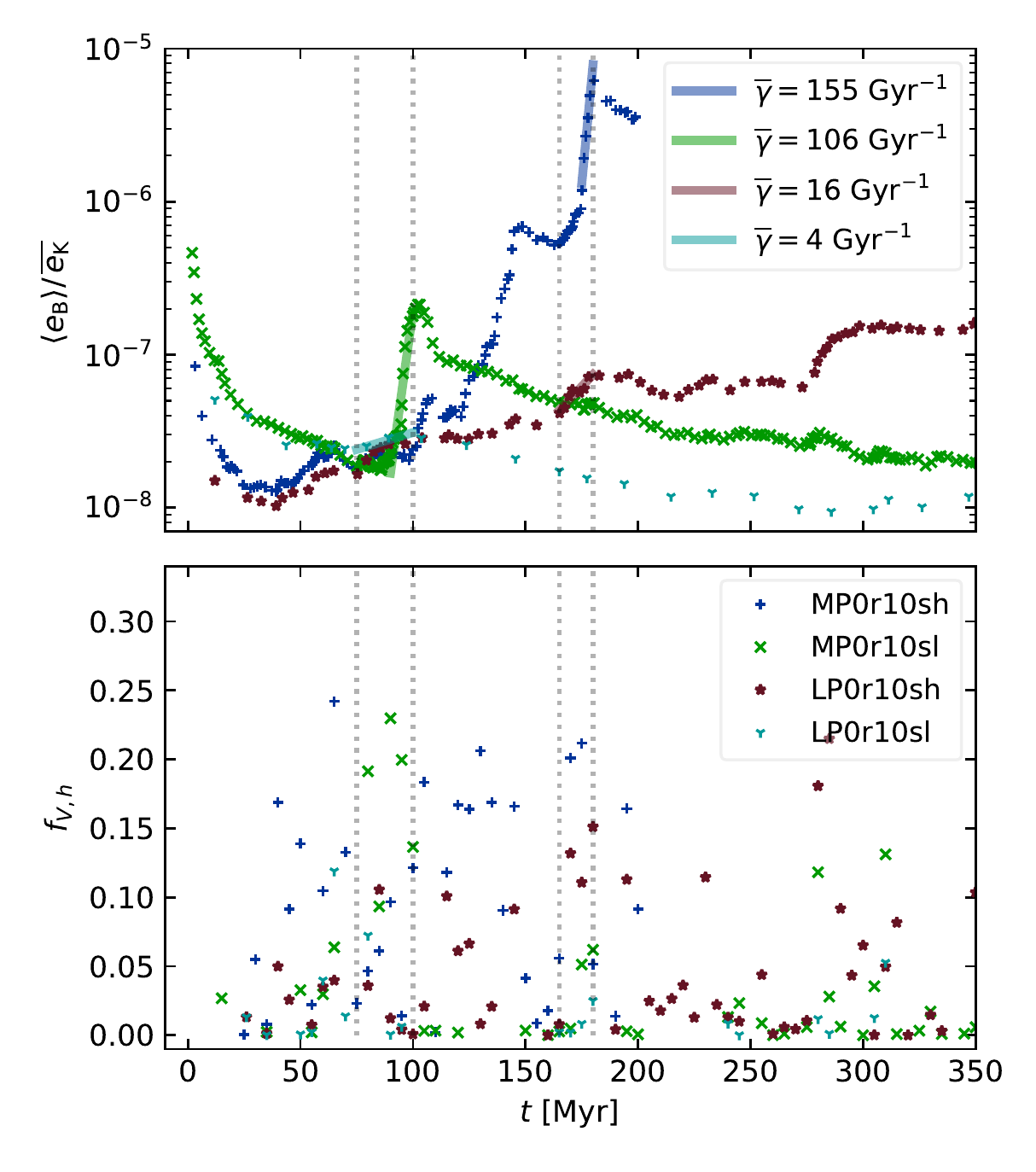}
  \begin{picture}(0,0)(0,0)
    \put(-115,282){{\sf\bf{(a)}}}
    \put(-115,152){{\sf\bf{(b)}}}
  \end{picture}
\caption{
	{\em (a)} Magnetic energy density with fits for growth rate $\barg$ as
	listed in the legend.  The fits for the $\delta x=4\pc$ models span the
	whole subintervals bounded by the vertical dotted lines, while the fits
	for $\delta x = 2\pc$ begin later, at $t=90$ and 175 Myr for low and
	high $\dot\sigma$, respectively.  Panel {\em (b)} for the same models,
	as listed in its legend for both panels, shows the evolution of
	fractional volume $\fH$ of the hot gas.
\label{fig:snrate} }
\end{figure}
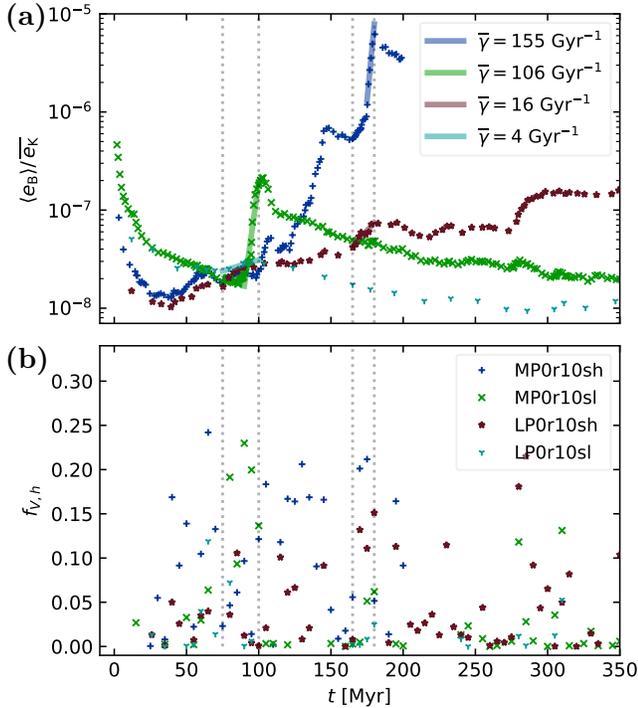

The strength of the SSD in SN-driven turbulence has been found to be sensitive
to the SN rate, $\dot\sigma$ \citep{BKMM04,GMKS21}.  \citet{BKMM04} used an SN
rate of $8\leq\dot\sigma\leq40\SNr$.  Because of their confinement within the
periodic boundaries, these models were dominated by thermal runaway, as
described by \citet{KO15}, in which the fractional volume of hot gas reaches
$\fH>0.9$.  Therefore, SSD has a steady exponential growth, because the ISM has
effectively become a homogeneous, single phase medium.  Indeed, in similar
experiments by \citet{GMKS21}, thermal runaway was easily excited at high
resolution ($\delta x\leq 1\pc$) even with $\dot\sigma=\SNr$.  At low
resolution ($\delta x = 4\pc$), with consequentially higher numerical
diffusion, and thus unphysically strong cooling, thermal runaway was still
excited for $\dot\sigma=10\SNr$.  Rapid steady acceleration of the SSD occurs,
which is excluded here with lower $\dot\sigma$. 

In Figure~\ref{fig:snrate}(a) we compare evolution of $\eB$ for
$\dot\sigma=0.2\SNr$ and $1\SNr$ at $\delta x=2\pc$ and 4 pc, and (b) the
fractional volume of the hot gas $\fH$.  The higher rate $\dot\sigma$ produces
more hot gas and sustains the dynamo, more so for $\delta x=2\pc$ {\em (blue)}
than $\delta x=4\pc$ {\em (brown)}.  For lower $\dot\sigma$, $\eB$ decays, but
growth occurs briefly in the subinterval between the first two vertical lines,
while $\fH$ increases, particularly for $\delta x=2\pc$ {\em (green)}.
Least-squares fits for $\barg$ are shown for this subinterval {\em (green and
cyan)}.  Higher $\fH$ occurs for higher $\dot\sigma$ with fits for $\barg$ {\em
(blue and red)} indicated in the second subinterval.  Another boost in SSD
occurs for $\delta x=4\pc$ after 270~megayears when $\fH$ is at an even higher
peak.  In general for $\delta x=4\pc$ there is less hot gas than for $\delta
x=2\pc$ at each given SN rate, because of higher numerical diffusion and
cooling.

Mean kinetic energy density is not strongly affected by the SN rate
\citep[][see their Figure 3{[d]}]{GMKS21}, so the change in SSD growth rate
would appear to be due to the increased heating of the ISM at higher
$\dot\sigma$.  Figure~\ref{fig:snrate} demonstrates that this is indeed the
case.  High SSD growth rate $\barg$ coincides with high $\fH$ for hot gas
within each model and between models, higher $\barg$ occurs in models that
support higher $\fH$.

\begin{figure}
\centering
\includegraphics[trim=0.0cm 0.0cm 0.0cm 0.0cm,clip=true,width=0.49\textwidth]{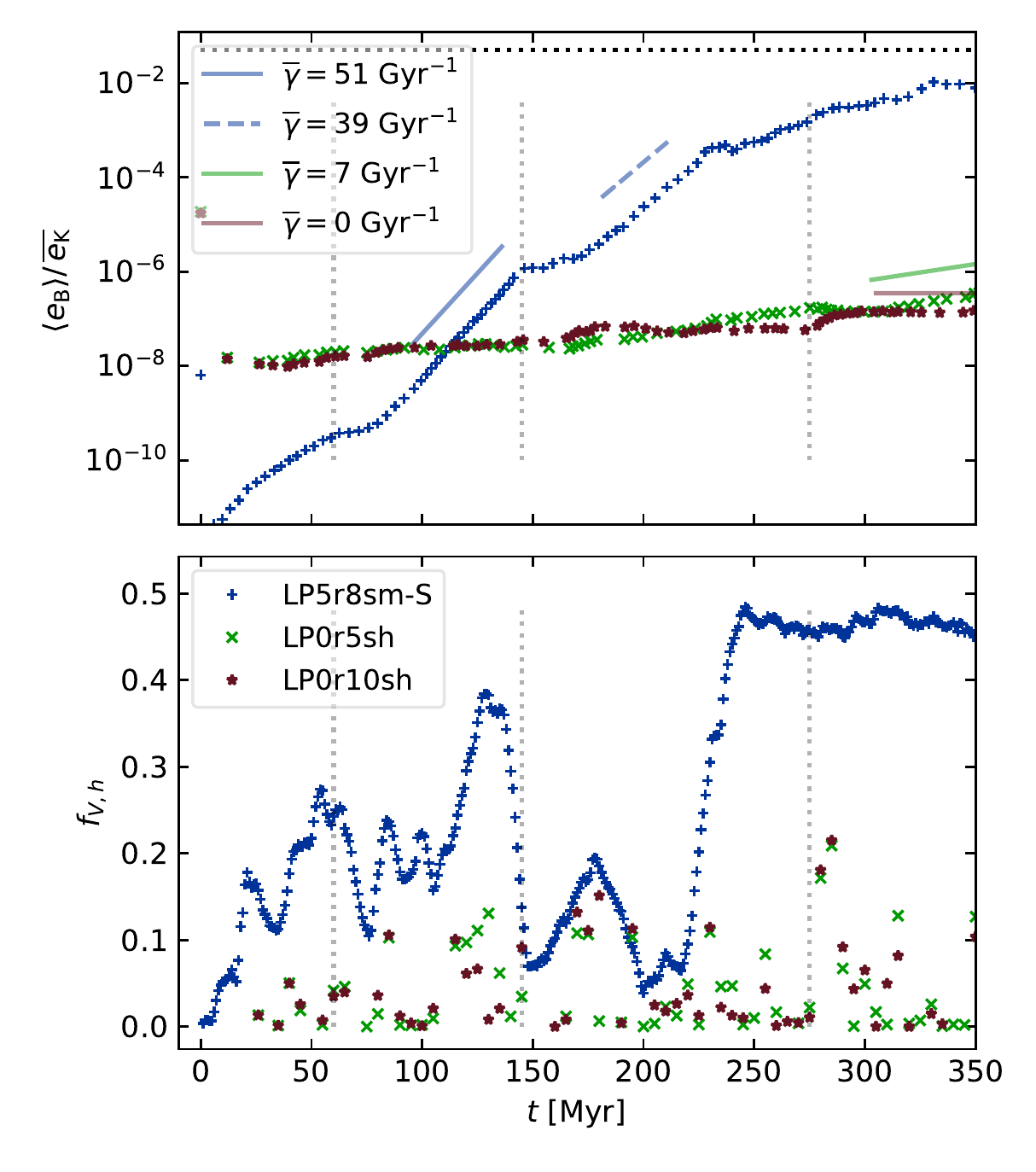}
  \begin{picture}(0,0)(0,0)
    \put(-115,282){{\sf\bf{(a)}}}
    \put(-115,152){{\sf\bf{(b)}}}
  \end{picture}
\caption{
	Magnetic energy density {\em (a)} with sample fits of growth rate
$\barg$ as indicated in the legend.  Panel {\em (b)} shows the evolution of
fractional volume $\fH$ of the hot gas for the same models, as listed in its
legend, including the stratified model.
	\label{fig:strat}
}
\end{figure}
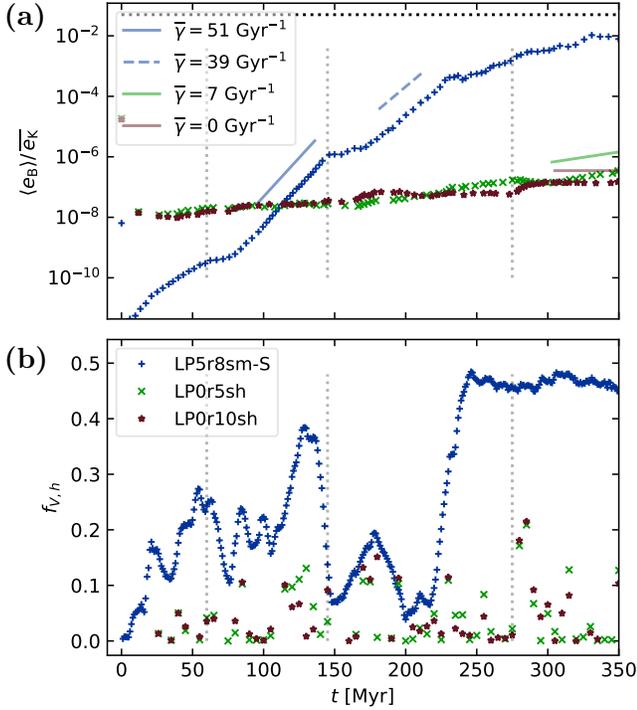

\subsection{Stratification}

Anticipating future examination of the LSD, in Model \lVmS\ we consider how
stratification affects the SSD.  To exclude the LSD, we omit large-scale
rotation and shear.  Inhomogeneities due to non-periodic boundaries vertically
might still cause an LSD effect, so any residual mean magnetic field volume and
horizontal component average is subtracted at each timestep.  In contrast to
the periodic domains, the open vertical boundary enables gas flow and magnetic
helicity flux to and from the disk.

In Figure~\ref{fig:strat}(a) we plot $\eB$ for this model with
$\dot\sigma=0.5\SNr$ alongside models with $\dot\sigma=\SNr$.  Given that
$\barg$ always increases with $\dot\sigma$ in otherwise equivalent models, it
is evident that the substantially increased growth rate in model \lVmS\ is due
to its stratification.  We also see from Figure~\ref{fig:strat}(b) that the
stratified Model \lVmS\ has the largest fractional volume of hot gas, although
none of the low resolution periodic models have large $\fH$ (see also
Figure~\ref{fig:snrate}]b]). 

From the summary statistics in Figure~\ref{fig:summary-vort}(b), we find for
$\delta x=4\pc$ that the absolute growth $\tilde\Gamma$ of magnetic energy is
fastest in the cold gas, but occurs mostly in the warm gas due to its higher
fractional volume.  In contrast the stratification supports high $\barg$ and
high $\fH$ for the hot gas, depicted in
Figures~\ref{fig:summary-vort}--\ref{fig:summary-urot} by a large orange plus
symbol.

As $\barg$ increases, the absolute growth rate $\tilde\Gamma$ in the hot gas
reduces, as shown in Figure~\ref{fig:summary-vort}(b) (left to right), for
$\delta x=4\pc$, While for $\delta x\geq2\pc$ it increases. In the cold gas
$\tilde\Gamma$ increases for $\delta x=4\pc$ and reduces for higher resolution.
At $\delta x=4\pc$, however, $\barg$ does not vary over time as much as for
higher resolution. At all stages $\tilde\Gamma$ in the warm gas is five to ten
times larger in the stratified model than the periodic models.

Vorticity in all phases is much lower (Figure~\ref{fig:summary-vort}) at
$\delta x=4\pc$, but of these it is highest in the stratified model. The values
of \Ms, $\Rm$, and velocities (Figs.~\ref{fig:summary-Ms} and
\ref{fig:summary-urot}), are also highest in the stratified model.  Higher
velocities, and vorticity, in all phases are possible with stratification,
because the gas can expand into lower pressure diffuse gas away from the
midplane.

\section{Discussion and conclusions}\label{sec:conc}

\subsection{Differences with isothermal dynamo}

Consensus based on studies of isothermal MHD turbulence predicts that $\Rm_{\rm
crit}$, the critical Rm above which SSD can be excited, reduces as Pm increases
above $\Pm\gtrsim0.1$ with an asymptotic limit of order $\Rm_{\rm crit}\sim20$
for $\Pm\gg1$ \citep{SBK02,Schober12,SBSW20}. Less directly relevant to our
context of ISM turbulence, for low Pm there is a range $0.1>\Pm>0.01$ where
$\Rm_{\rm crit}$ has a maximum \citep{SHBCMM05, ISCM07, B11}, and then for
$\Pm<0.01$ $\Rm_{\rm crit}$ again begins to drop \citep{WKGR22}.  Once excited,
the rate of growth of SSD also increases, up to some asymptote, for higher Pm.
In these isothermal systems, it has also been demonstrated that as \Ms\
increases it becomes more difficult to excite SSD \citep[$\Rm\gtrsim60$ for
$\Pm\geq5$][]{Haugen:2004M} and growth rates are correspondingly reduced
\citep{FCSBKS11,FSBS14}.  Hence, according to these findings, SSD can be
excited in our models, as this threshold value is easily exceeded.

Superficially, these conclusions might appear  irrelevant to an SSD in
multiphase SN-driven turbulence with respect to the averages across all phases.
The response of models with respect to Pm is inconsistent. Higher SN rates with
correspondingly higher compressive forcing appear to excite SSD more easily and
with larger growth rate.  SSD does not appear to follow a dominant eigenmode,
establishing a distinct exponential amplification until saturation.

However, detailed analysis reveals that what is inconsistent in this more
realistic ISM is the distribution of temperatures, gas densities, and flow
properties throughout the thermal phases and over time.  Determination of Pm,
Rm, Re, and \Ms\ must be specific to each region of the ISM.  It is likely that
the variation we see by position in these properties occurs also in isothermal
simulations, but that the statistics of these remain quite consistent
throughout the domain and duration. In a multiphase system, on the other hand,
these statistics vary by orders of magnitude as a function of position and
time.

The breakdown of the models by phase and overall SSD growth rate shows that the
insights from isothermal SSD experiments do apply, but contrary trends, such as
increasing growth in SSD with increasing \Ms, seen in
Figure~\ref{fig:summary-Ms}(a), are due to the SSD supportive counter trends,
such as increasing velocity (with \Ms). The expected effect of \Ms\ on growth
rate is well demonstrated in the anticorrelation going from hot to cold phase.
Conversely, in Figure~\ref{fig:summary-Ms}(b), Rm seems anticorrelated with SSD
growth rate in the hot phase, indicating that SSD is more easily excited as the
hot filling fraction and typical temperature of the hot gas increase, even for
lower Rm.  The expected trend of positive correlation between Rm and SSD growth
rate is evident in the differences between phases.

\subsection{Growth rates and strength of SSD}

In their two-phase simulations, \citet{SF22} determined that the magnetic
energy in their models grows more slowly than in isothermal models with the
same forcing and equivalent mean Mach numbers \Ms$\sim1$ for the warm gas
and 5 for the cold gas.  Their SSD saturates after around 400~megayears for
solenoidal forcing and 800 megayears for compressive forcing, for a grid
resolution comparable to Model \uOl.  For this and all our 1 pc resolution
periodic models, the SSD saturates within 40--60 megayears with mean
\Ms $\sim1$, 0.6 and 0.25 in the cold, warm and hot gas, respectively.
However, in all phases \Ms\ spans around two orders of magnitude over a
large proportion of the volume, and outliers extend even further.  The
variance is much larger than in the two-phase simulations.  The faster SSD
in our models may, therefore, be less sensitive to the mean \Ms, as growth
occurs predominantly in low Mach number, high vorticity regions,
particularly in the hot gas. 

\rev{The SSD persists longest in the cold phase and next longest in the warm
phase (Fig.~\ref{fig:Pm}[c]). We speculate that an explanation for the growth
rates differing between phases in three-phase and not in two-phase simulations
is that high vorticity generated in the hot gas transfers to warm then cold
regions, but with a time delay that is significantly longer than from the warm
to cold phase. As a result, we find faster growth in all phases.}

Estimates for the post inflationary seed magnetic field are in the range
$10^{-65}$--$10^{-20}\upmu$G \citep{GR01,Kandu16}.  With an exponent of
$\barg=500\Gyr^{-1}$ this could be amplified to sub-equipartition saturation of
around $1\upmu$G within 170 to 590~Myr.  Given these results consistently yield
saturation of the SSD significantly below equipartition with the kinetic energy
density \citep[see also][]{SSFBK15}, this does not exclude LSD action being
sufficient to amplify even the lowest estimates of post-inflation magnetic
field to the values observed in high redshift galaxies.

\subsection{Interpretation of Rm}

It is perhaps surprising, and inconsistent with current understanding of
dynamo, to see in Figure~\ref{fig:Rmhist} locations with $\Rm\leq1$, well below
isothermal predictions of $\Rm_{\rm crit}$, coincident with growth rate
$\tilde\Gamma>10^3\Gyr^{-1}$ in cold gas.  These are outliers in a highly
inhomogeneous multiphase structure, in which there is phase mixing and energy
transfer within and between phases.  The bulk of the simulation data
corresponds to high Rm and high $\tilde\Gamma$ with the positive correlation
reflected in the increasing log mean from the cold to hot phase.

To examine whether  the high apparent growth rates at low Rm can
be explained without SSD, let us approximate the effect of a blast wave
following Sedov-Taylor \citep{ Sedov59,Taylor50} travelling at $150\kms$, as
would occur in an SN remnant with radius near 30 pc in a $10^4$~K medium of
uniform gas number density $1\cmcube$.  From Equation~\eqref{eq:tildeG} with
the normalized quantity
$\hat{\vect{B}}=\vect{B}\langle{\mu_0e_B}\rangle^{-1/2}$ we can write 
\begin{align}
\tilde\Gamma & = -\hat{B}^2
\nabla\cdot\vect{u}+\hat{\vect{B}}\cdot(\hat{\vect{B}}\cdot\nabla)\vect{u}
-\\  \nonumber
 & - \hat{\vect{B}}\cdot(\vect{u}\cdot\nabla)\hat{\vect{B}} + 
\eta\hat{\vect{B}}\cdot\nabla^2\hat{\vect{B}}.
\end{align}
Considering the first term with a shock front traveling at $150\kms$ into warm
gas with sound speed around $12\kms$ resolved in our model over three cells,
$-\nabla\cdot\vect{u}\approx 138\kms(3\pc)^{-1}\approx 4.6 \times 10^4
\Gyr^{-1}$. If the local magnetic field is even twice the volume averaged
strength, this would yield $\tilde\Gamma\gtrsim10^5\Gyr^{-1}$.  The effect of
the other terms and nonadiabatic cooling would adjust this estimate up or down,
but nevertheless it remains entirely consistent with the rare high values of
$\Gamma$ and $\tilde\Gamma$ in regions of low Rm in
Figure~\ref{fig:Rmhist}.

For a magnetic field perpendicular to the blast wave 
\begin{equation} \Rm\sim\frac{|\partial_\perp u_\perp
B_\parallel|}{|\eta\partial^2_\perp B_\parallel|}, 
\end{equation}
reaching a maximum at the shock centre, where the first derivative approaches
infinity and the second derivative vanishes, but near zero either side of the
shock where the inverse is true.  Both can coexist with the compression
$\tilde\Gamma$ estimated above, explaining these low Rm outliers in the joint
histograms.

Conventionally, a single characteristic Rm for a domain is related to $\barg$,
associated with the single exponential growth rate of the dominant eigenmode of
the SSD.  The analysis of \citet{GOBSC17} found that the Lorentz forces could
induce weak nonlocal energy transfers, but that magnetic and kinetic energy
transfers in both directions are predominantly highly localised in $k$-space,
including for isothermal compressible MHD.  Hence, the differences in
characteristic length scales between the phases would tend to yield Rm
supportive of very different dominant dynamo modes.  Thus, in the multiphase
ISM the SSD growth rate $\barg$ represents a superposition of varying dynamo
modes, in which the dominant modes can switch from one phase to another over
time due to heating, cooling and phase mixing. 

The mean $\Gamma$ of the instantaneous localised growth rates collected while
$\gamma(t)$ is highest have similar magnitude to $\barg$ as fitted in each
simulation, supporting the concept of a superposition.  Sufficiently high mean
$\Gamma$ or $\tilde\Gamma$ correlated with a supercritical mean Rm would
dominate the overall growth rates irrespective of the outliers.

\subsection{SSD dependence on vorticity}

We considered the properties in the simulations of velocity, Mach number, fluid
and magnetic Reynolds numbers, Prandtl number, kinetic and magnetic helicity,
and kinetic energy to identify how they relate to the relative and absolute
growth rates of magnetic energy $\Gamma$ and $\tilde\Gamma$.  By far the
clearest correlation to growth rate was in the enstrophy, or norm squared
vorticity.

The dependence of growth rate on velocity shows a weaker dependence, which is
very similar to that of the Helmholtz decomposed flow displayed in
Figure~\ref{fig:summary-urot}, and even less correlation is seen in the kinetic
energy density.  The latter indicates that it is the magnitude of the flow
rather than the kinetic energy density that is important to the SSD efficiency.

We break down the statistics into epochs of various rates of volume averaged
growth $\barg$. Periods of higher $\barg$ have higher velocities and vorticity.
From the stronger correlation between vorticity and growth, we conclude that
the vorticity of the flow drives the SSD.  Despite the compressive structure of
the SN forcing there is, as expected, a high fraction of rotational flow
present, as identified from the Helmholtz decomposition\rev{. The excess of
rotational to compressive flow is indicated by the compressive ratio at less
than 40\% and even below 10\% in the hot gas.}  The efficiency of vortex
generation is primarily driven in the ISM by the strong baroclinicity, the
cross of orthogonal gradients of pressure and density, at the SN shock
interactions \citep[see][]{KGVS18}.  Typically the magnitude of the velocity
and the \rev{proportion of} enstrophy increase with temperature. \rev{Fitting
data from Figure~\ref{fig:summary-vort}(a) and for the high resolution data
only from Figure~\ref{fig:summary-urot}(c), \rev{we find that the local growth
rate} $\langle\Gamma\rangle\propto\langle|\omega|\rangle$ and
$\langle\Gamma\rangle\propto\langle\Rey\rangle$.}

A concern might exist that vorticity could be artificially generated, as has
been identified at the intersection of adaptive grids of mixed mesh refinement
\citep{PM01}, or due to the use of locally varying diffusive schemes
\citep{RKGAR10} at lower resolution.  Diffusivity is independent of position in
our models, except for its use in resolving shocks.  Inspection of the energy
spectra \citep[as in][and here in Figure~\ref{fig:power} of the Appendix]{GMKS21}
confirms in all cases that energy peaks are far from the Nyquist frequency and
with very low energy near theses scales.  Hence, any transfer of energy from
the flow to the magnetic field must occur at length scales independent of the
shock resolution over a few cells.

The study of vortex generation by \citet{KGVS18} was a purely hydrodynamical
study, although stratified, with similar levels of vortex generation. It had no
hyperdiffusion.  Given the relatively low magnetic energy here, even after
saturation of the SSD, we can also exclude the generation of vorticity being
due to the magnetic field.

\subsection{SSD dependence on temperature}\label{sec:temp}

We have shown that SSD activity increases when the fractional volume of the hot
gas increases, and we have shown that the magnetic energy density in the hot
gas grows the most rapidly during these epochs. At other times, though, the
magnetic field does not grow as quickly in the hot gas as in the warm gas.  The
vorticity in the hot gas is also particularly high during these rapid bursts of
SSD.  So, both the volume and the stirring of the hot gas increase, combining
to drive strong dynamo action for a limited period.

In \citet{KMG22}, we use model \uOl\ to examine dust-processing due to SN blast
waves. The progress of individual remnants are followed in detail for isolated
explosions in a dense turbulent region or a diffuse turbulent region.  The
effects on the magnetic field inside each remnant is instructive for the
understanding of SSD in hot gas (see their Figure~1). In the case of the
explosion in a dense region, the magnetic field is evacuated from within the
remnant and is swept along by the the blast wave for up to 1~megayear.  In the case
of the diffuse region, the surrounding regions of strong magnetic field are
pushed away by the blast wave, but inside the remnant the magnetic field grows,
with the interior field strength orders of magnitude stronger at 1~megayear than at
the time of the explosion.  The SSD is enhanced in the hot gas where the
interaction of the blast wave with the inhomogeneous density structure excites
strong, interacting reverse shocks that can persist to late times in the
diffuse medium.  In dense regions, on the other hand, the reverse shocks from
the ambient ISM are weak relative to the blast wave and then quickly slow in
the relatively high gas density remnant interior.

This is consistent with our finding that the SSD is more sensitive to the
vorticity than the kinetic energy, as the diffuse remnant contains relatively
low kinetic energy.  In our stratified model \lVmS\ the SSD is more active,
with the ISM typically more diffuse.  The SSD accelerates when the hot gas
develops more solenoidal turbulence through SN remnant interactions and mergers
\citep[such as emerging superbubbles in the local and nearby galactic
disks][]{BOMAHHM18,SK21}.  Simulated superbubbles with magnetic fields
\citep[e.g.,][]{FMZ91,Stil_2009} have tended to apply uniform ambient fields
and smooth ambient gas, such that amplification of the magnetic fields are
concentrated around the outer shell due to compression, as in the case of the
dense region. Superbubbles in turbulent MHD have been modelled
\citep[e.g.,][]{Korpi:1999b,deAvillez:2005,BA06,GEZR08,Gent:2013b,Gent:2013a},
but the local structure of the magnetic field not analysed in such detail.

\subsection{Summary of results}
We conclude that
the SSD in the SN-driven multiphase turbulence of the ISM:
\begin{enumerate}[a.]
	\item	
    has an average growth rate \rev{linearly} correlated with the average
		\rev{vorticity} of the flow, which is efficiently generated by
		baroclinic effects due to SN blast wave interactions;
	\item  correspondingly has an average growth rate well correlated
		with the average magnitude of the rotational flow velocity, \rev{as well as the compressible and total velocities}  \rev{and the fluid Reynolds number}, 
		\rev{and anticorrelates with the compressive ratio};
	\item	grows as a superposition of varying dynamo modes in each phase and within
		different regions of each phase;
	\item  grows intermittently, predominantly during periods when
		the hot gas, with low sonic Mach number \Ms, has strong
		enstrophy and large volume filling fraction;
	\item   is thus most efficient in hot gas produced when
		supernovae explode in diffuse regions, where
		interacting reverse shocks can have high velocities;
	\item is more efficient in a stratified disk, where diffuse gas away
		from the disk is more easily heated and supports high
		velocities with low \Ms;
	\item and already appears to approach asymptotic solutions at values of
		$\Rm$, $\Rey$ and $\Pm$ significantly below the actual values
		estimated to apply in the ISM, although higher resolution will
		be required to confirm this point (see Appendix).

\end{enumerate}
Prior to the onset of SSD the magnetic field strength aligns with the gas
density, consistent with turbulent mixing, and following saturation of the SSD
the field redistributes back into alignment with the gas concentration.  The
hot gas saturates significantly below equipartition, and even below the the ISM
average of around 5\% of equipartition with kinetic energy density, while the
cold gas saturates at close to the equipartition energy.

\begin{acknowledgments}
 F.A.G. and M.J.K.-L. acknowledge support from the Academy of Finland
ReSoLVE Centre of Excellence (grant 307411), the Ministry of Education and
Culture Global Programme USA Pilot 9758121 and the ERC under the EU's
Horizon 2020 research and innovation programme (Project UniSDyn, grant
818665) and generous computational resources from CSC -- IT Center for
Science, Finland, under Grand Challenge GDYNS Project 2001062.
M.-M.M.L.
was partly supported by US NSF grant AST18-15461.
\rev{We acknowledge the constructive criticism of the anonymous referee, which
	helped us to improve the paper.}
\end{acknowledgments}

\software{Pencil Code\footnote{\href{
https://github.com/pencil-code}{https://github.com/pencil-code}}
 \citep{brandenburg2002,Pencil-JOSS}}
\clearpage
\bibliographystyle{aasjournal}
\bibliography{refs}

\appendix

\section{Effective Prandtl number}\label{sec:ap}
\begin{figure}[h]
\centering
\includegraphics[trim=0.4cm 0.4cm 0.35cm 0.3cm,clip=true,width=0.48\textwidth]{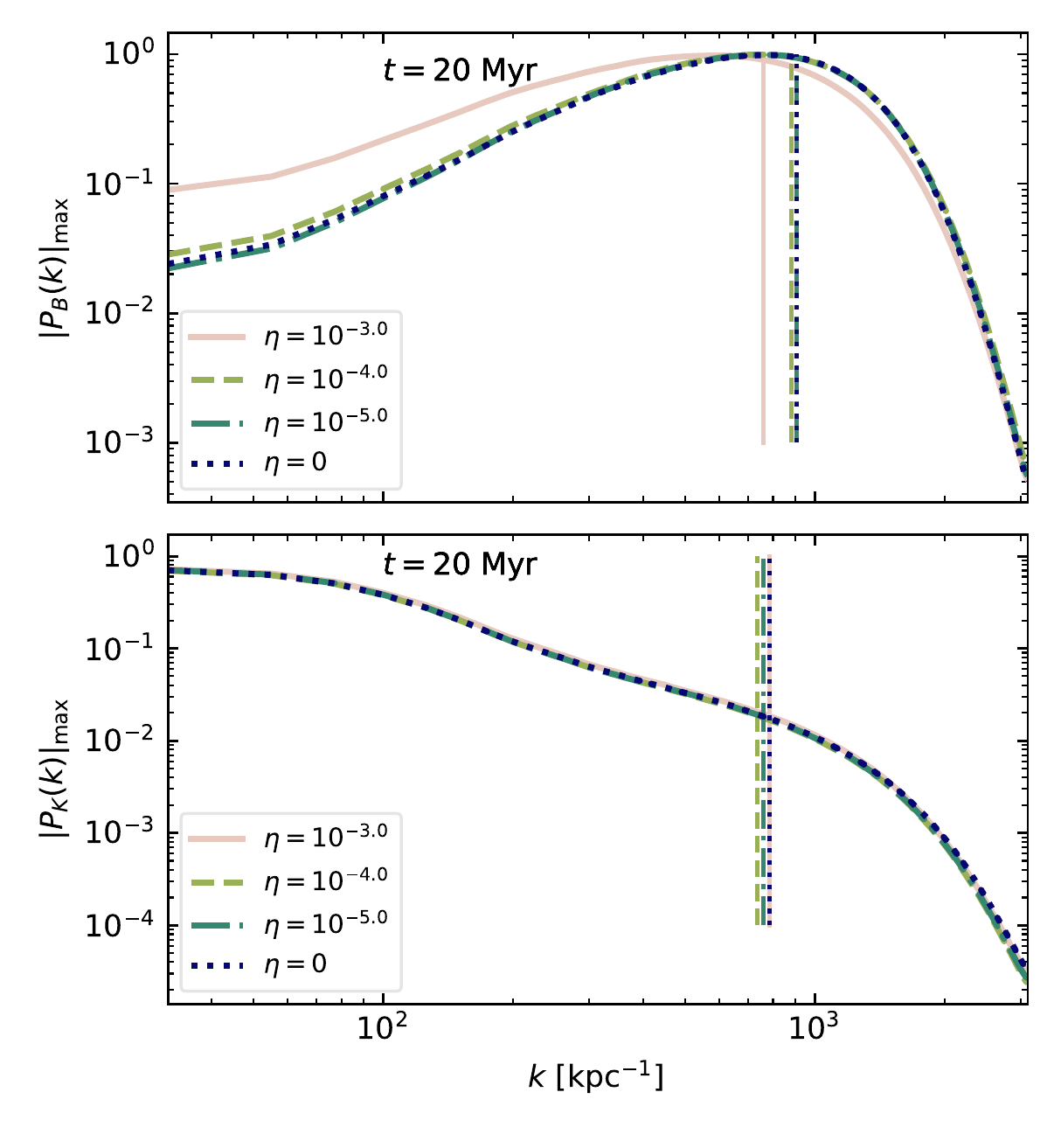}
\includegraphics[trim=0.4cm 0.4cm 0.35cm 0.3cm,clip=true,width=0.48\textwidth]{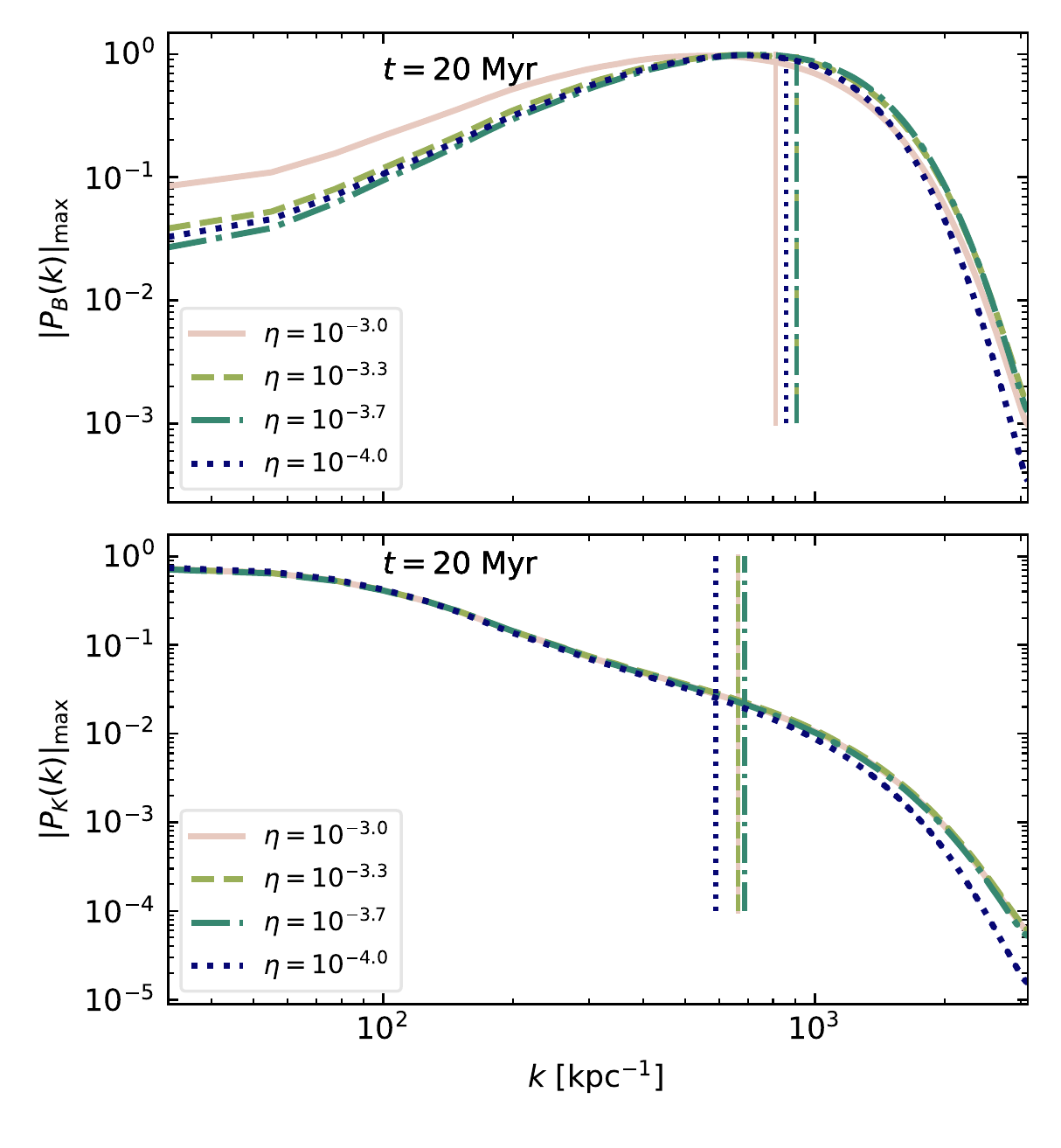}
  \begin{picture}(0,0)(0,0)
    \put(-500,256){{\sf\bf{(a)}}}
    \put(-250,256){{\sf\bf{(b)}}}
  \end{picture}
\caption{
	Snapshots at times indicated of magnetic and kinetic energy spectra for
$\delta x=1\pc$ periodic models with viscosity (a) $\nu=10^{-3}$ and (b)
$\nu=0$, with resistivity $\eta$ as listed in the legends.  $k$ representing
the beginning of the dissipative range are identified in each spectrum by the
vertical lines of matching color and style.  $k_\eta$ ($k_\nu$) is the lowest
$k$ for which the gradient of the spectrum is below (above) $-7.5 \times 10^{-5}$ for the
magnetic (kinetic) spectra, in each case corresponding to the increase in the
rate of decay of the spectrum. 
\label{fig:power}
}
\end{figure}
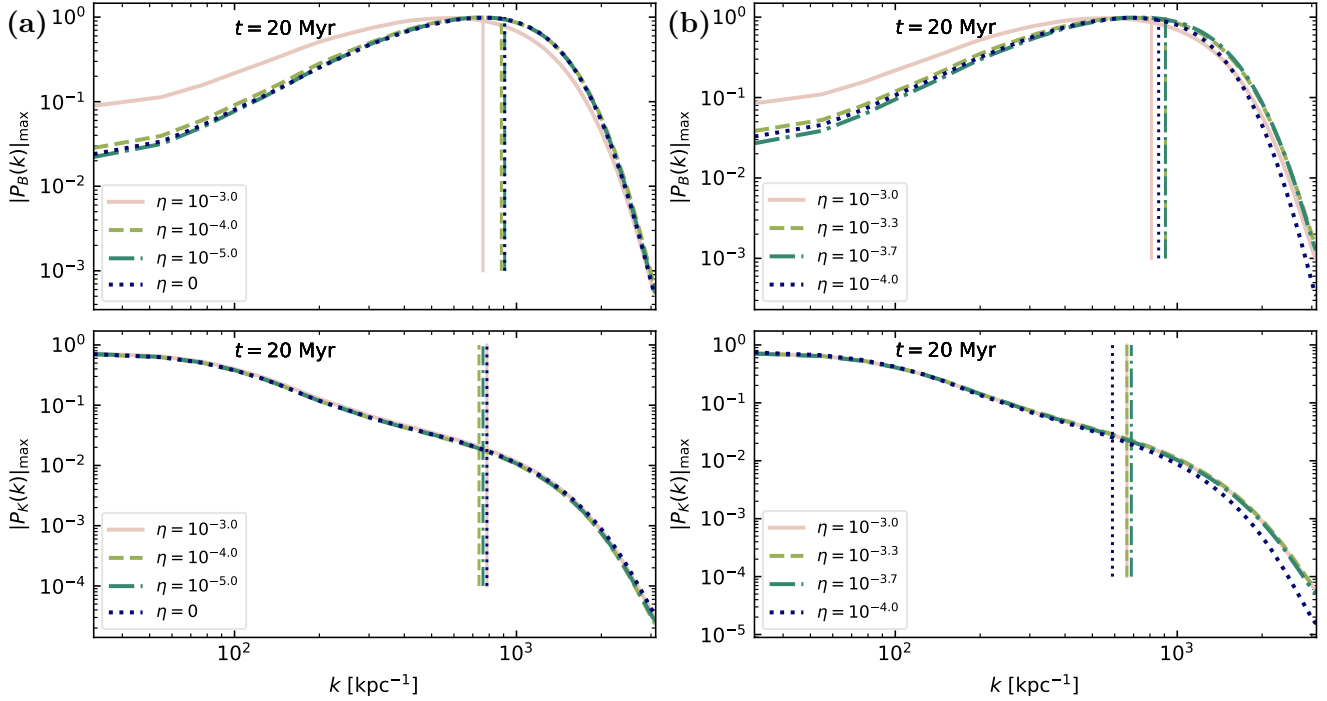

Examples of the magnetic and kinetic energy spectra are displayed in
Figure~\ref{fig:power} to confirm that the nominal $\Pm = \nu / \eta$ are
reflected in the effective magnetic Prandtl number 
\begin{equation}\label{eq:Pmeff}
	\Pm_{\rm eff} = \frac{k_\eta}{k_\nu},
\end{equation}
in which $k_\eta$ and $k_\nu$ are the wavenumbers where the dissipative range
begins. At the time indicated, early in the kinematic stage, the spectra are
averaged over snapshots spanning $\pm2\Myr$  and then Gaussian smoothed with
$\sigma=2\delta k$.  To aid comparison the spectra are all normalised by their
maximum. 

The wavenumbers and their ratios $\Pm_{\rm eff}$ are listed in
Table~\ref{tab:Pmeff} for each parameter $\nu$ and $\eta$ in the models.  These
include some models not listed in Table~\ref{tab:models}, but included in
\citet{GMKS21}.  $k_\nu$ is determined as the lowest wavenumber $k$ after the
intertial range for which the log gradient of $\log(P_K)<-5/3$ and for
$k_\eta$, where the log gradient of $\log(P_B)<-2/3$.  In the case of $\nu=0$
for $\eta\leq10^{-4}$, the numerical diffusivities of hyperdiffusion and shock
diffusion appear to dominate, with $\Pm_{\rm eff}\approx1.2$. For
$\eta=10^{-3}$ this is sufficient to obtain $\Pm_{\rm eff}<1$, which should
nominally be the case for any $\eta<\nu$.

For $\nu=10^{-3}$ we obtain $\Pm_{\rm eff}>1$ in all cases, due to the higher
effective numerical viscosity than resistivity. Nonetheless, the difference
between $\eta=10^{-3}$ and $\eta\leq0.005$ is sufficient to explain the
differences, between the $\Pm=1$ model and $\Pm=5$. However, we should be
cautious about concluding whether the SSD is converging above Pm=5 until we
increase resolution or reduce the effective numerical diffusivities.

\begin{table*}[h]
	\centering
	\caption{
		Estimated effective magnetic Prandtl numbers $\Pm_{\rm eff}$ for models with various physical
		viscosity $\nu$ and resistivity $\eta$.
		\label{tab:Pmeff}}
\begin{tabular}{>{\centering\arraybackslash}p{2.8cm}>{\centering\arraybackslash}p{2.8cm}>{\centering\arraybackslash}p{2.8cm}>{\centering\arraybackslash}p{2.8cm}>{\centering\arraybackslash}p{2.8cm}>{\centering\arraybackslash}p{2.8cm}}
\hline\hline
 $\nu$ & $\eta$ & $k_\nu$ & $k_\eta$ & $\Pm$   & $\Pm_{\rm eff}$ \\\hline
 0     &  1e-03 & 785.4 & 760.9      & 0       & 0.97 \\ 
 0     &  1e-04 & 736.3 & 883.6      & 0       & 1.20 \\ 
 0     &  1e-05 & 760.9 & 908.1      & 0       & 1.19 \\ 
 0     &  0e+00 & 785.4 & 908.1      & \nodata & 1.16 \\ 
 1e-3  &  1e-03 & 662.7 & 809.9      & 1       & 1.22 \\ 
 1e-3  &  5e-04 & 662.7 & 908.1      & 2       & 1.37 \\ 
 1e-3  &  2e-04 & 687.2 & 908.1      & 5       & 1.32 \\ 
 1e-3  &  1e-04 & 589.0 & 859.0      & 10      & 1.46 \\\hline 
\end{tabular}
\end{table*}

\end{document}